\documentclass[useAMS,usenatbib,usegraphicx]{mn2e}

\def\kms{\,km\,s$^{-1}$}
\def\teff{$T_{\rm eff}$}

\title[Spectropolarimetry of cool white dwarfs]{A study of high proper-motion
white dwarfs\\
I. Spectropolarimetry of a cool hydrogen-rich sample\thanks{Based on 
observations made with ESO telescopes at the La Silla Paranal Observatory 
under programme IDs 80.D-0521, 82.D-0521, 83.D-0750 and 84.D-0862.}}
\author[Kawka and Vennes]{Ad\'ela Kawka\thanks{E-mail:
kawka@sunstel.asu.cas.cz (AK); vennes@sunstel.asu. cas.cz (SV)} and 
St\'ephane Vennes\footnotemark[2] \\
Astronomick\'y \'ustav AV \v{C}R, Fri\v{c}ova 298,CZ-251 65 Ond\v{r}ejov,
Czech Republic}

\begin{document}

\date{Accepted; Received}

\pagerange{\pageref{firstpage}--\pageref{lastpage}} \pubyear{2012}

\maketitle

\label{firstpage}

\begin{abstract}
We conducted a spectropolarimetic survey of 58 high proper-motion white dwarfs which achieved uncertainties
of $\ga 2$ kG in the H$\alpha$ line and $\ga 5$ kG in the upper Balmer line series. 
The survey aimed at detecting low magnetic fields ($\la 100$ kG) and helped identify 
the new magnetic white dwarfs NLTT~2219, with a longitudinal field $B_{\rm l} = -97$ kG, and
NLTT~10480 ($B_{\rm l}=-212$ kG). Also, we report the possible identification of a 
very low-field white dwarf with $B_{\rm l} = -4.6$ kG. The observations show that $\approx5$\% of 
white dwarfs harbour low fields ($\sim10$ to $\sim10^2$~kG) and that increased survey sensitivity may help uncover
several new magnetic white dwarfs with fields below $\sim$1 kG.
A series of observations of the high field white dwarf NLTT~12758 revealed
changes in polarity occurring within an hour possibly associated to an inclined, fast rotating dipole.
Also, the relative strength of the $\pi$ and $\sigma$ components in NLTT~12758 possibly revealed
the effect of a field concentration (``spot''), or, most likely, the presence of
a non-magnetic white dwarf companion.
Similar observations of NLTT~13015 also showed possible polarity variations, but without a clear indication
of the timescale.
The survey data also proved useful in constraining the chemical composition, 
age and kinematics of a sample of cool white dwarfs as well as in constraining the incidence of double
degenerates. 

\end{abstract}

\begin{keywords}
stars: fundamental parameters -- stars: magnetic field -- white dwarfs
\end{keywords}

\section{Introduction}
 
\begin{figure*}
\centering
\includegraphics[width=1.0\columnwidth]{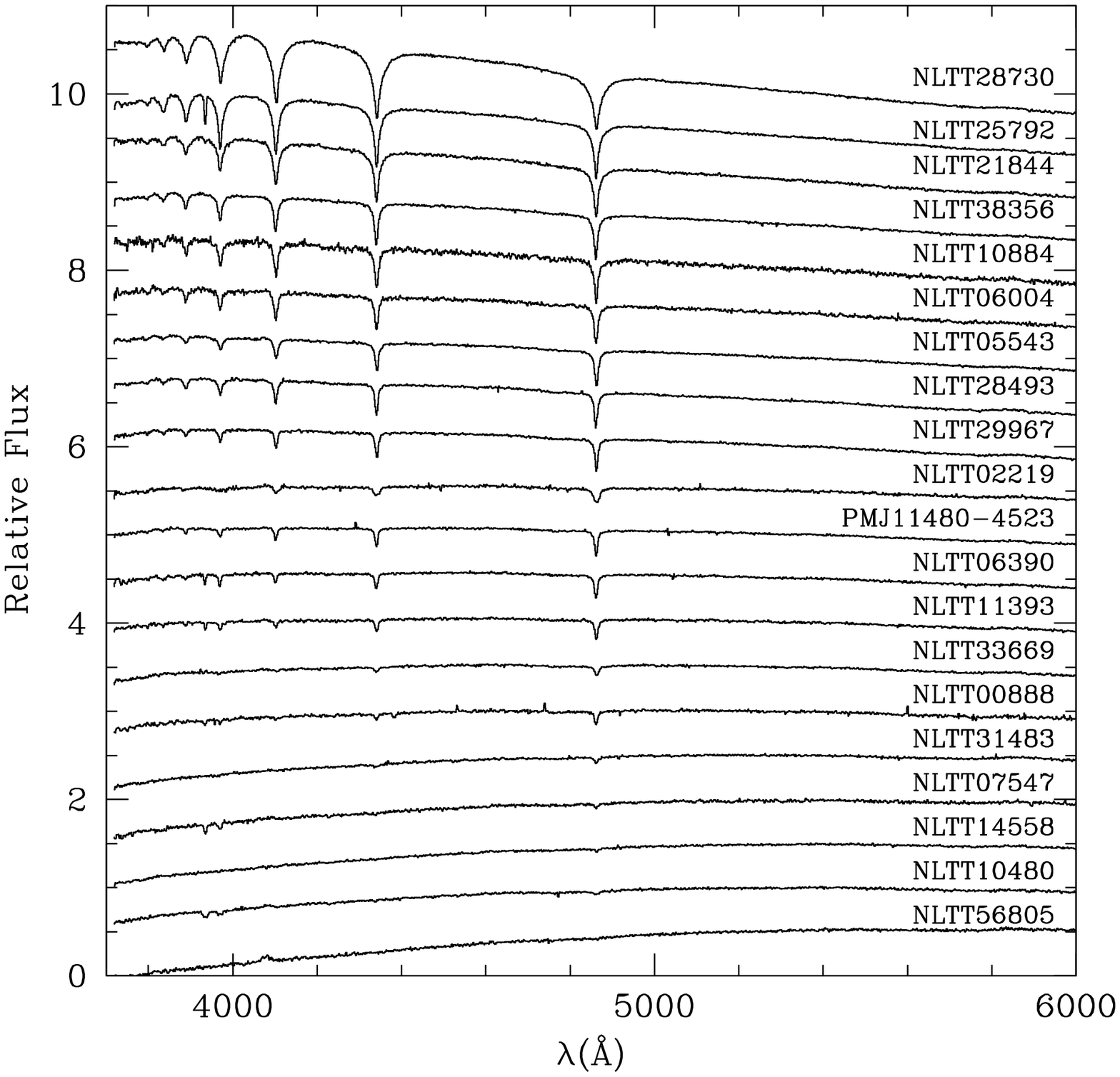}%
\includegraphics[width=1.0\columnwidth]{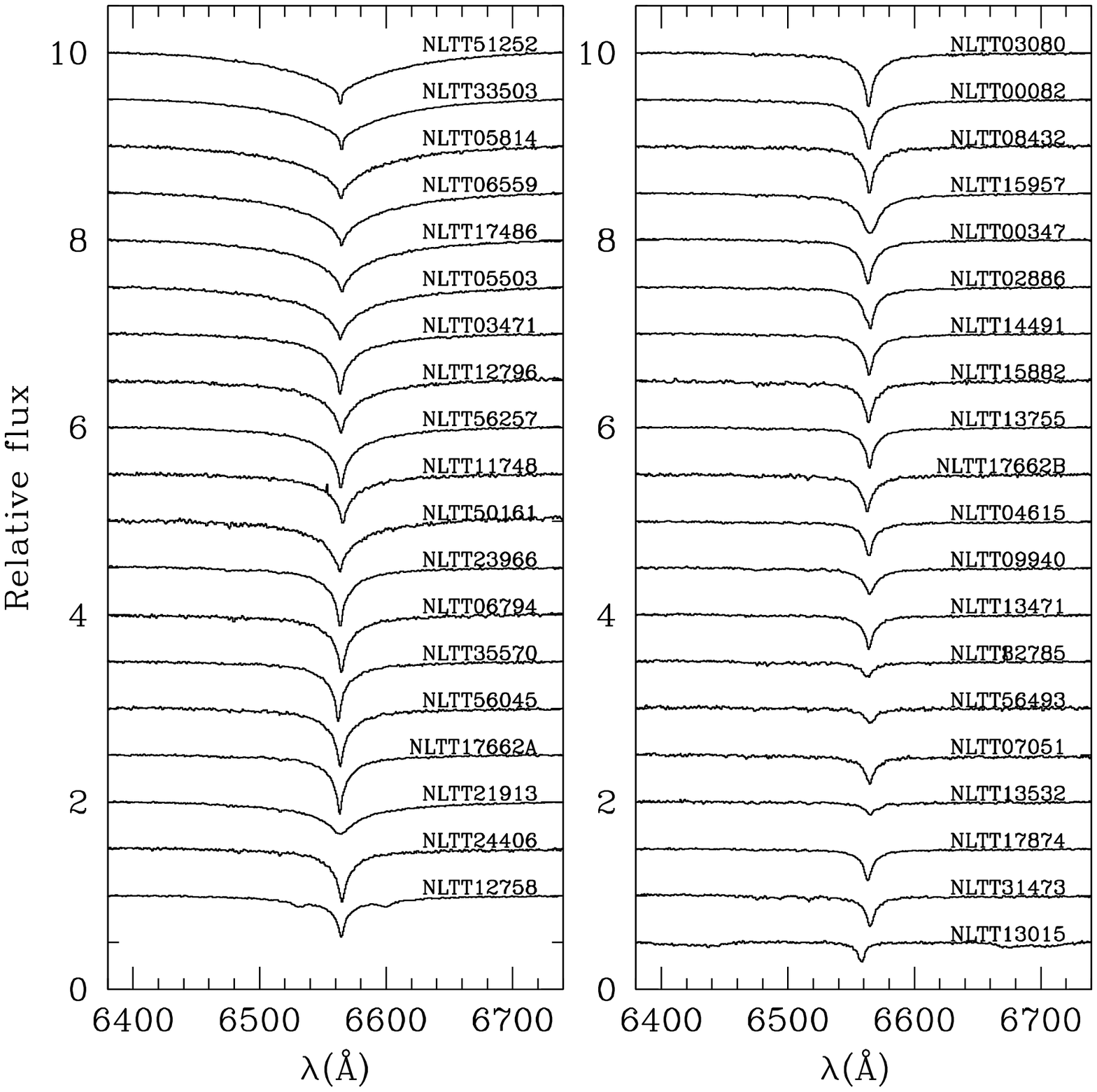}%
\caption{VLT/FORS spectra of DA white dwarfs ordered by increasing temperatures 
(bottom to top and right to left): 20 spectra were obtained (left panel) using 
grism 600B and covering the upper
Balmer lines, and (right panel) 39 spectra were obtained using grism 1200R+93 and 
covering H$\alpha$. Several objects show Ca~K (NLTT~888, 6390, 7547, 10480, 11393, and 25792) in their spectra 
with a few more showing Ca{\sc i}$\lambda$4226 (NLTT~7547, 10480, and 25792) and 
are classified as DAZ white dwarfs. Zeeman line splitting is immediately apparent in the H$\alpha$ spectra
of NLTT~12758 and 13015.} 
\label{fig_DA}
\end{figure*}

Magnetic fields in white dwarfs show great diversity in strength and structure.
Field strengths from a few kilogauss \citep[e.g., 40 Eri B or LTT~9857,][]{fab2000,azn2004}
to nearly 10$^9$ G \citep[e.g., J0317$-$855,][]{bar1995,fer1997} delineate the current range of detection, although 
higher and lower fields cannot, in principle, be ruled out.
Low-dispersion spectropolarimetric surveys successfully achieved detection limits below
a kilogauss \citep{azn2004} or of the order of a few kG \citep{sch1994,kaw2007}, while echelle spectroscopy achieved limits of several kilogauss 
with Zeeman splitting measurements in line cores \citep{koe1998}. The spectra of high-field white dwarfs are very complex and
detailed modelling requires application of model atoms in high-fields \citep{mar1984,jor1992}. Hydrogen and helium
atomic data are available for such studies \citep{kem1974,gar1974}. 

Stellar rotation, by exposing a variable line-of-sight projection of the field vector, often reveals
a more complex field structure than a simple dipole \citep{mar1984} as found in the cases
of the fast-rotating ($P\approx12$ min) white dwarf J0317$-$855 \citep[see][]{bur1999,ven2003x}
or the ``spotted'' magnetic white dwarf WD~1953$-$011 with a rotation period of 1.4 day \citep{max2000,bri2005,val2008}. 
Apart from the common hypothesis of a fossil origin,
magnetic fields may be generated in later stages as first proposed by \citet{tou2008}, first, during a common-envelope (CE) phase of binary evolution \citep[see also][]{pot2010,nor2011} or in a double degenerate merger \citep[see also][]{gar2012}.

\citet{kaw2004x} examined the distribution of field strength in the Ap star population and concluded
that a direct link between Ap stars and high field magnetic white dwarfs ($B\ga 10^7$ G) may exist assuming magnetic flux conservation, but that low-field white dwarfs are without known progenitors.
However, \citet{wic2005} found a progenitor deficit for high-field white dwarfs and
postulated that a large fraction of massive main-sequence stars ($\ga 4.5\,M_\odot$) may
harbour weak fields (10-100~G).

We present a survey of 58 high proper-motion white dwarfs conducted with the FOcal Reducer and low dispersion Spectrograph (FORS)
at the European Southern Observatory (ESO). The observations, presented in Section 2, provided us with intensity and circular polarization spectra
enabling high-sensitivity measurements of the surface-averaged ($B_{\rm S}$) and longitudinal ($B_{\rm l}$) magnetic fields (Section 3). 
In particular, we examine the sample properties (Section 3.1), our new low-field detections (Section 3.2) and instances of variable high-fields (Section 3.3),
and the sample kinematical properties (Section 3.4) as well as other interesting aspects of this survey (Section 3.5).
We summarize and discuss some implications of our results in Section 4.

\section{Observations}

\begin{figure*}
\centering
\includegraphics[width=1.0\columnwidth]{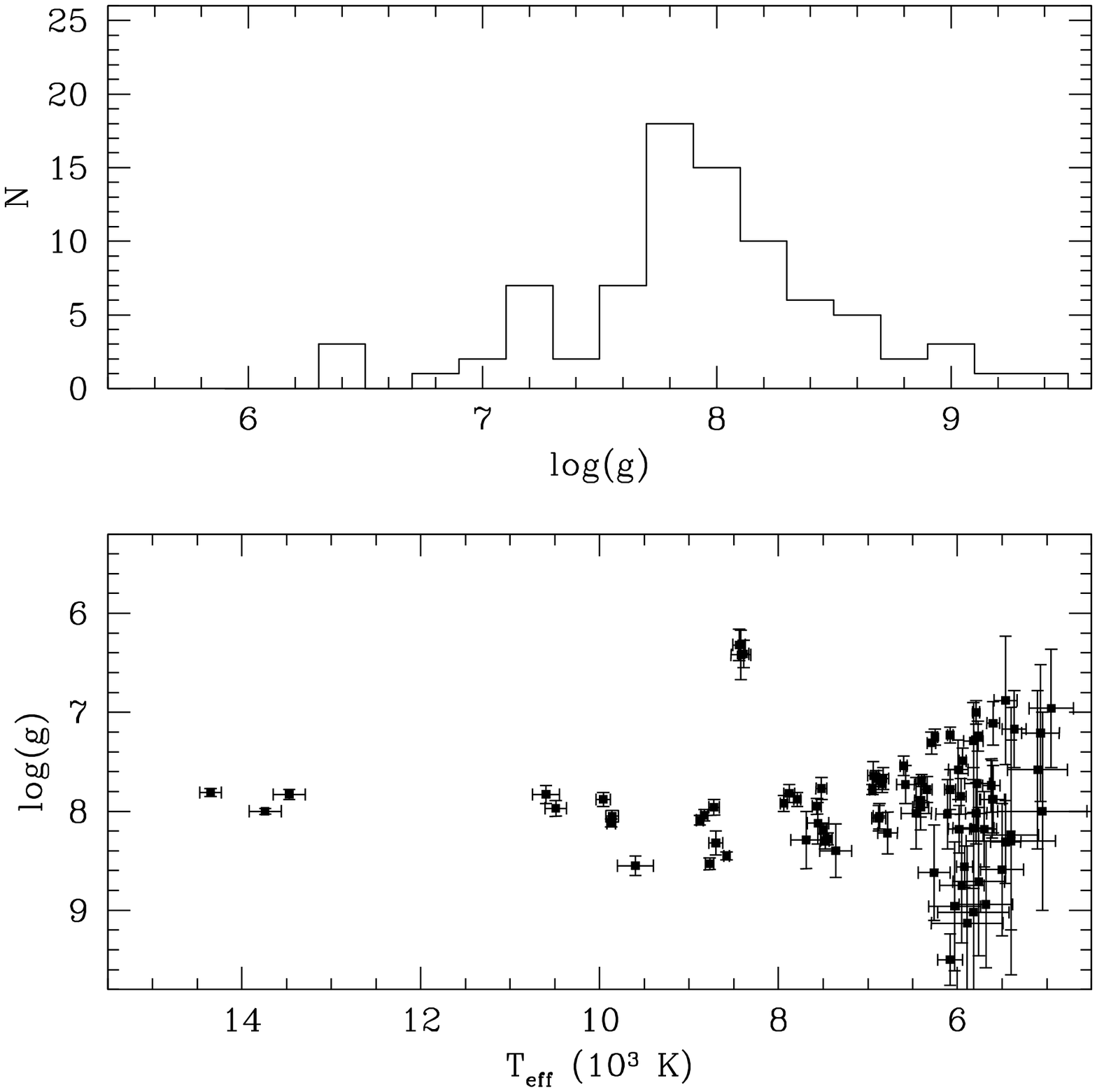}
\includegraphics[width=1.0\columnwidth]{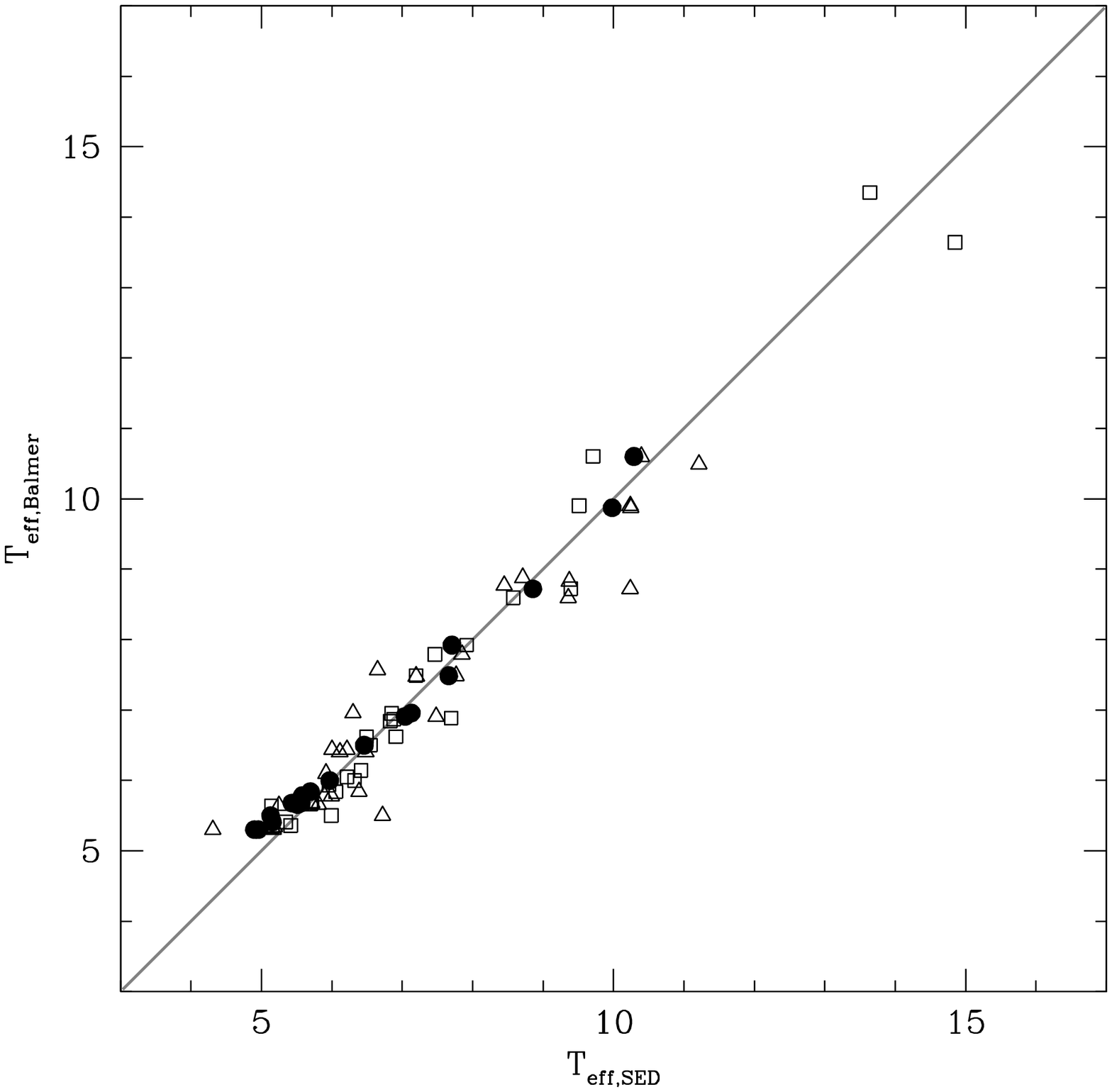}
\caption{(Bottom left panel) set of 83 surface gravity and effective temperature measurements of the sample of
56 stars that exclude the two high-field DAP white dwarfs. The measurement errors increase with decreasing effective temperature. The low-gravity star near
\teff$\,=8500$ K is NLTT~11748. (Top left panel) the distribution of surface gravity measurements shows high-
and low-gravity tails that are mostly accounted for by large errors in surface gravity measurements. (Right panel) Adopted effective temperatures
versus effective temperatures determined using colours described in Appendix A: SDSS $ugriz$ (full circles), $V-J$ (open squares), and $R-J$ (open triangles).}
\label{fig_tg}
\end{figure*}

We obtained two sets of observations using the FORS spectrograph attached to
the 8m telescopes at Paranal Observatory. The first data set was obtained
with the FORS1 attached to UT2. We used the 600 lines per mm grism centred at
4650 \AA\ (Grism 600B). The slit width was set to 1.0 arcsec to obtain a
spectral dispersion of 1.51\AA\ per pixel and a resolution of $\sim 6$\AA. The
second set of observations was obtained with the FORS2 attached to UT1. For
these observations we used the 1200 lines per mm grism centred at 6500 \AA\
(Grism 1200R+93) providing a spectral dispersion of 0.73 \AA\ per pixel. The
slit width was set to 1.0 arcsec to obtain a spectral resolution of 3.0 \AA.
Both sets of spectra were obtained in the spectropolarimetric mode where the
observing sequence consisted of an exposure with the Wollaston prism rotated
to $-45^\circ$ immediately followed by an exposure with the Wollaston prism
rotated to $+45^\circ$.

Additional spectra of the programme stars were obtained at ESO using the
New Technology Telescope (NTT) and at Cerro Tololo Inter-American Observatory 
(CTIO) and Kitt Peak National Observatory (KPNO). We also collected spectra 
from the Sloan Digital Sky Survey (SDSS) Data Release 8, and recycled our 
published spectra \citep{kaw2006}. The complete data set will be described in 
a forthcoming publication and, here, we employ key results of the spectroscopic 
analysis of these complementary spectra. In summary, the spectra obtained with 
the NTT have a resolution of $\sim 14$ \AA\ and a spectral range of
3700 - 7400 \AA, the spectra obtained with the CTIO 4m telescope have a
resolution of $\sim 8$ \AA\ with a spectral range of 3700 - 7480 \AA, and
the spectra obtained with the KPNO 4m telescope have a resolution of
$\sim 5.5$ \AA\ and a spectral range of 3660 - 6790 \AA. The SDSS spectra 
cover a range from 3800 to 9200 \AA\ and have a resolution of $\sim 3$ \AA.

The data were reduced using standard IRAF procedures. The stars observed
during P80 (FORS1) were flux calibrated using the standard stars Feige 110, 
EG 21 and EG 274.

Table~\ref{tbl_obs} lists the selected targets. Figure~\ref{fig_DA}
shows the intensity spectra obtained during the two FORS programmes. 
Some of the blue spectra showed heavy element lines (Ca~H\&K) characteristic
of polluted DA white dwarfs (DAZ), while the red spectra only showed H$\alpha$ 
with the exception of the spectrum of NLTT~11748: the strong, saturated Na~D 
doublet ($W(D2)=220\pm20$m\AA, $W(D2)=210\pm20$m\AA) at a heliocentric velocity 
of $v=20\pm3$\kms revealed a high neutral gas (Na{\sc i}) column density in the interstellar medium (ISM) toward the star. 
The Na~D doublet was not detected toward any other objects from our sample.
The star NLTT~11748 is located toward a known high-density line-of-sight \citep{ver2001,wel2010}
and the scaled ISM H{\sc i} column density ($10^{21}$ cm$^{-2}$) correlates well with the
relatively large extinction coefficient $E_{B-V}=0.1$ measured by \citet{kaw2009}.

Observations of the DC white dwarfs NLTT~8525, NLTT~8904, NLTT~11051, NLTT~20389,
and PM~J04356$-$6105, of the DZ white dwarfs NLTT~57760 and PM~J08186$-$3110, and
of the DQ white dwarfs NLTT~8733 and NLTT~14553 will be presented in a forthcoming
publication.
Two targets that were subsequently reclassified are not listed: NLTT~18393 (M3.5) and
NLTT~40020 (G5V). Finally, we also excluded
one observation of NLTT~17486 obtained on UT 2007 October 31 and all observations of NLTT~40636 because 
of pointing errors. The coordinates and proper-motion measurements are from \citet{sal2003}.

The circular polarization spectra were obtained using two successive 
exposures obtained with the Wollaston prism rotated to the angle 
$\theta = -45^\circ$ and $+45^\circ$ and were calculated following:
\begin{equation}
v = V/I = \frac{1}{2}\Big[\Big(\frac{f_o - f_e}{f_o + f_e}\Big)_{\theta = 45^\circ} - \Big(\frac{f_o - f_e}{f_o + f_e}\Big)_{\theta = -45^\circ}\Big]
\end{equation}
where $f_o$ and $f_e$ are the fluxes from the ordinary and extra-ordinary
beams, respectively. The direction of polarization of these beams are 
perpendicular to each other. This method is used to eliminate possible biases
to first order \citep{bag2002}.

\subsection{Imaging and photometry}

We calibrated the acquisition images and converted the count rates into broad-band magnitudes.
The procedure is described in Appendix A and the magnitudes are listed in Table~\ref{tbl_mag} along
with published $V$ magnitudes and Two Micron All Sky Survey (2MASS) $J$ magnitudes.  
A close examination of the acquisition images of NLTT~17662 revealed the the presence of a close (sep.$\approx 1.6$ arcsecond, $P.A.=340^\circ$) companion
to the white dwarf approximately 1 mag fainter. A comparison with the first-epoch Digitized Sky Survey (DSS) image proves that the 
faint companion is in common proper-motion with the bright star. Fortunately, both stars were acquired during the spectroscopic observations
and we were able to analyse both FORS spectra. 
Similarly, our acquisition image of NLTT~56045 also reveal the presence of a close
common proper-motion companion 1.1 mag fainter in white light (sep.$\approx 2.2$ arcsecond, $P.A.=290^\circ$). The faint component was also acquired during 
spectroscopic observations but no spectra were extracted.

Finally, we cross-correlated our sample of stars with the 
SDSS Data Release 8 and obtained $ugriz$ photometry for 18 objects. Only one of 
these had a SDSS spectrum. Table~\ref{tbl_sdss} lists the available $ugriz$ 
photometry. A correction of $-0.04$ mag was applied to the tabulated $u$ magnitude
prior to our analysis \citep[see, e.g.,][]{kle2004}.

\begin{table*}
\begin{minipage}{\textwidth}
\caption{VLT FORS observations.\label{tbl_obs}}
\centering
\renewcommand{\footnoterule}{\vspace*{-15pt}}
\begin{tabular}{llcccccc}
\hline\hline
Name & Alternate Names & RA (2000) & Dec (2000) & $V_p$ \footnote{Photographic magnitudes from \cite{sal2003}.}
 & $\mu$ & UT Date & $N$ \footnote{Number of complete spectropolarimetric sequences.} \\
     &                 &           &            & (mag) & ($\arcsec$\,yr$^{-1}$) & & \\
\hline
NLTT   82       & LP644-30, PHL2595, G158-128 & 00 04 10.34 & $-$03 40 08.54 & 16.73 & 0.237 & 25 Oct 2009 & 2 \\
NLTT  347       & LP764-69, G158-134          & 00 08 21.70 & $-$14 35 02.18 & 16.86 & 0.169 & 25 Oct 2009 & 2 \\
NLTT  888       & LHS5002, LP644-81           & 00 17 37.72 & $-$05 16 46.42 & 17.82 & 0.475 & 15 Nov 2007 & 1 \\
                &                             &             &                &       &       & 12 Dec 2007 & 1 \\
NLTT 2219       & LP645-70, PB8467            & 00 40 56.18 & $-$08 09 08.82 & 17.19 & 0.426 & 14 Dec 2007 & 2 \\
NLTT 2886       & LP882-92                    & 00 52 08.94 & $-$30 36 39.67 & 16.48 & 0.196 & 15 Oct 2009 & 2 \\
NLTT 3080       & G270-98, EG6, LP706-65      & 00 55 50.34 & $-$11 27 31.50 & 15.36 & 0.439 & 25 Nov 2009 & 1 \\
                &                             &             &                &       &       & 02 Dec 2009 & 1 \\
                &                             &             &                &       &       & 21 Dec 2009 & 1 \\
NLTT 3471       & PHL3287, G270-123, LP646-83 & 01 03 19.63 & $-$03 24 59.76 & 16.60 & 0.178 & 25 Oct 2009 & 1 \\
                &                             &             &                &       &       & 26 Oct 2009 & 1 \\
NLTT 4615       & LP587-53                    & 01 23 14.71 & $-$02 09 27.00 & 17.13 & 0.246 & 26 Oct 2009 & 2 \\
NLTT 5503       & G271-117, LP648-27, PHL1086 & 01 38 52.58 & $-$03 56 50.17 & 16.43 & 0.187 & 02 Dec 2009 & 1 \\
                &                             &             &                &       &       & 03 Dec 2009 & 1 \\
NLTT 5543       & LHS1274, LP939-114          & 01 39 14.38 & $-$33 49 03.87 & 17.34 & 0.595 & 08 Oct 2007 & 2 \\
NLTT 5814       & GD1394, MCT0142$-$3026      & 01 44 36.27 & $-$30 11 22.38 & 16.22 & 0.189 & 15 Oct 2009 & 1 \\
NLTT 6004  & G274-95, LP884-35, HE0145$-$2726 & 01 47 43.93 & $-$27 11 36.89 & 15.84 & 0.320 & 03 Nov 2007 & 1 \\
NLTT 6390       & LP884-58                    & 01 54 05.36 & $-$30 34 32.41 & 17.08 & 0.489 & 01 Nov 2007 & 1 \\
NLTT 6559       & LP649-6, KUV01552-0703      & 01 57 41.91 & $-$06 48 46.87 & 16.57 & 0.261 & 06 Jan 2010 & 1 \\
                &                             &             &                &       &       & 25 Jan 2010 & 1 \\
NLTT 6794       & G274-137, LP884-82          & 02 01 41.26 & $-$26 50 38.26 & 16.87 & 0.255 & 16 Oct 2009 & 1 \\
                &                             &             &                &       &       & 25 Jan 2010 & 1 \\
NLTT 7051       & LP885-22                    & 02 07 02.33 & $-$30 23 32.42 & 16.18 & 0.266 & 15 Oct 2009 & 1 \\
NLTT 7547       & LP649-83                    & 02 17 19.63 & $-$06 56 28.86 & 17.89 & 0.391 & 13 Dec 2007 & 2 \\
NLTT 8432       & LP830-13                    & 02 35 21.69 & $-$22 51 21.96 & 17.02 & 0.278 & 30 Nov 2009 & 1 \\
                &                             &             &                &       &       & 01 Feb 2010 & 1 \\
NLTT 9940       & LP651-74                    & 03 07 13.91 & $-$07 15 06.23 & 17.23 & 0.491 & 24 Nov 2009 & 2 \\
NLTT10480       & LHS5070, LP887-66           & 03 17 12.08 & $-$29 11 34.33 & 17.30 & 0.496 & 01 Nov 2007 & 1 \\
NLTT10884       & LP772-61                    & 03 25 15.50 & $-$17 22 27.77 & 16.46 & 0.320 & 27 Oct 2007 & 1 \\
NLTT11393       & LP832-30, WT1384            & 03 36 34.09 & $-$22 15 24.05 & 17.21 & 0.346 & 03 Nov 2007 & 1 \\
NLTT11748       & LP413-40                    & 03 45 16.83 & $+$17 48 08.71 & 16.66 & 0.295 & 22 Oct 2009 & 3 \\
                &                             &             &                &       &       & 23 Oct 2009 & 2 \\
NLTT12758       & G160-51                     & 04 12 26.33 & $-$11 17 47.33 & 15.46 & 0.285 & 23 Oct 2009 & 3 \\
                &                             &             &                &       &       & 24 Nov 2009 & 2 \\
NLTT12796       & GD58                        & 04 13 38.67 & $-$08 01 28.88 & 16.37 & 0.202 & 03 Dec 2009 & 1 \\
                &                             &             &                &       &       & 15 Jan 2010 & 1 \\
NLTT13015       & LP714-52                    & 04 19 21.10 & $-$09 34 29.57 & 17.50 & 0.182 & 22 Oct 2009 & 1 \\
                &                             &             &                &       &       & 10 Nov 2009 & 1 \\
                &                             &             &                &       &       & 24 Nov 2009 & 1 \\
NLTT13471       & LP775-28                    & 04 32 00.35 & $-$19 20 49.85 & 16.97 & 0.263 & 09 Jan 2010 & 1 \\
                &                             &             &                &       &       & 25 Jan 2010 & 1 \\
                &                             &             &                &       &       & 31 Jan 2010 & 1 \\
NLTT13532       & LP890-39                    & 04 33 33.58 & $-$27 53 24.79 & 16.68 & 0.388 & 16 Oct 2009 & 1 \\
                &                             &             &                &       &       & 25 Jan 2010 & 1 \\
                &                             &             &                &       &       & 01 Feb 2010 & 1 \\
NLTT13755       & LP 775-37                   & 04 41 05.03 & $-$15 19 03.51 & 16.83 & 0.183 & 09 Dec 2009 & 2 \\
NLTT14491       & LP776-52                    & 05 05 32.43 & $-$17 31 37.41 & 17.13 & 0.314 & 24 Jan 2010 & 2 \\
NLTT14558       & LHS1739, LP777-3            & 05 08 36.36 & $-$15 23 13.06 & 17.69 & 0.599 & 31 Jan 2008 & 2 \\
NLTT15882       & LP659-7                     & 05 59 00.96 & $-$04 14 26.88 & 17.00 & 0.245 & 30 Jan 2010 & 2 \\
                &                             &             &                &       &       & 01 Feb 2010 & 2 \\
NLTT15957       & G105-4                      & 06 02 31.10 & $+$15 53 04.85 & 16.76 & 0.315 & 09 Dec 2009 & 2 \\
NLTT17486       & LHS1898, LP896-18           & 07 09 25.09 & $-$32 05 07.30 & 15.87 & 0.530 & 15 Oct 2009 & 1 \\
NLTT17662A/B    & G89-10, LTT17958, EG171     & 07 18 08.64 & $+$12 29 59.24 & 15.89 & 0.298 & 08 Dec 2009 & 1 \\
NLTT17874       & LP428-34                    & 07 27 04.15 & $+$14 34 40.22 & 16.96 & 0.195 & 12 Dec 2009 & 2 \\
NLTT21844       & G49-7, G41-39, LP427-59, SDSS & 09 28 40.22 & $+$18 41 14.42 & 15.92 & 0.302 & 01 Mar 2008 & 1 \\
NLTT21913A/B    & EC09273-1719, LP787-49      & 09 29 43.17 & $-$17 32 50.54 & 15.93 & 0.449 & 09 Dec 2009 & 1 \\
NLTT23966       & LP669-73                    & 10 18 33.18 & $-$04 42 28.58 & 16.76 & 0.168 & 23 Jan 2010 & 2 \\
\hline
\end{tabular}
\end{minipage}
\end{table*}

\begin{table*}
\begin{minipage}{\textwidth}
\contcaption{}
\centering
\renewcommand{\footnoterule}{\vspace*{-15pt}}
\begin{tabular}{llccccccc}
\hline\hline
Name & Alternate Names & RA (2000) & Dec (2000) & $V_p$ \footnote{Photographic magnitudes from \cite{sal2003}, except PMJ11480$-$4523 from \citet{lep2005}.} & $\mu$ & UT Date & $N$ \footnote{Number of complete spectropolarimetric sequences.} \\
     &                 &           &            & (mag) & ($\arcsec$\,yr$^{-1}$) & & \\
\hline
NLTT24406       & LP430-25, SDSS              & 10 27 47.79 & $+$19 28 24.53 & 17.10 & 0.383 & 03 Feb 2010 & 1 \\
                &                             &             &                &       &       & 09 Feb 2010 & 1 \\
                &                             &             &                &       &       & 14 Feb 2010 & 1 \\
NLTT25792       & EC10542-2236, LP849-31      & 10 56 38.63 & $-$22 52 55.96 & 15.64 & 0.302 & 07 Mar 2008 & 1 \\
NLTT28493       & LHS2455, G11-020            & 11 46 25.78 & $-$01 36 36.76 & 16.19 & 0.562 & 29 Mar 2008 & 1 \\
NLTT28730       & LP673-41                    & 11 50 33.33 & $-$06 36 17.06 & 16.56 & 0.323 & 28 Mar 2008 & 1 \\
                &                             &             &                &       &       & 30 Mar 2008 & 1 \\
NLTT29967       & LP674-29                    & 12 12 36.02 & $-$06 22 18.12 & 17.26 & 0.443 & 30 Mar 2008 & 1 \\
NLTT31473       & LHS339, LP853-15            & 12 40 24.15 & $-$23 17 43.80 & 16.68 & 1.114 & 31 Jan 2010 & 1 \\
                &                             &             &                &       &       & 01 Feb 2010 & 1 \\
                &                             &             &                &       &       & 09 Feb 2010 & 1 \\
                &                             &             &                &       &       & 05 Mar 2010 & 1 \\
                &                             &             &                &       &       & 06 Mar 2010 & 1 \\
NLTT31483       & LHS2601, LP435-447          & 12 40 30.51 & $+$18 07 28.96 & 17.42 & 0.595 & 31 Mar 2008 & 2 \\
NLTT32785       & LP676-58                    & 13 04 46.39 & $-$05 28 37.74 & 17.12 & 0.278 & 31 Jan 2010 & 1 \\
                &                             &             &                &       &       & 05 Mar 2010 & 1 \\
                &                             &             &                &       &       & 07 Mar 2010 & 1 \\
NLTT33503       & LP737-47, LHS271            & 13 16 43.60 & $-$15 35 58.45 & 15.07 & 0.710 & 05 Mar 2010 & 1 \\
NLTT33669       & LP854-50, WT2034            & 13 19 24.76 & $-$21 47 54.85 & 16.35 & 0.455 & 09 Mar 2008 & 1 \\
NLTT35570       & LP912-27                    & 13 53 47.16 & $-$27 38 54.24 & 16.75 & 0.198 & 05 Mar 2010 & 2 \\
NLTT38356       & LP914-20                    & 14 47 22.33 & $-$30 35 23.89 & 17.44 & 0.323 & 31 Mar 2008 & 2 \\
NLTT50161       & LP928-33                    & 20 56 08.33 & $-$27 17 23.64 & 17.19 & 0.171 & 29 Nov 2009 & 1 \\
NLTT51252       & LP873-45, EG502             & 21 26 26.30 & $-$22 43 53.33 & 15.59 & 0.207 & 28 Nov 2009 & 1 \\
NLTT56045A/B    & LP933-63                    & 23 10 46.39 & $-$29 48 30.64 & 16.52 & 0.271 & 21 Dec 2009 & 1 \\
NLTT56257       & LP822-40                    & 23 15 30.40 & $-$14 40 05.81 & 15.13 & 0.186 & 30 Nov 2009 & 1 \\
                &                             &             &                &       &       & 16 Dec 2009 & 1 \\
NLTT56493       & LP702-67                    & 23 19 35.43 & $-$02 29 00.96 & 16.89 & 0.238 & 16 Dec 2009 & 1 \\
NLTT56805       & LP522-46                    & 23 25 19.86 & $+$14 03 39.42 & 15.81 & 0.353 & 27 Oct 2007 & 1 \\
PMJ11480$-$4523 & WD1145$-$451                & 11 48 03.32 & $-$45 23 01.8  & 15.66 & 0.635 & 07 Mar 2008 & 1 \\
\hline
\end{tabular}
\end{minipage}
\end{table*}

\section{Analysis}

\begin{figure*}
\centering
\includegraphics[width=\columnwidth]{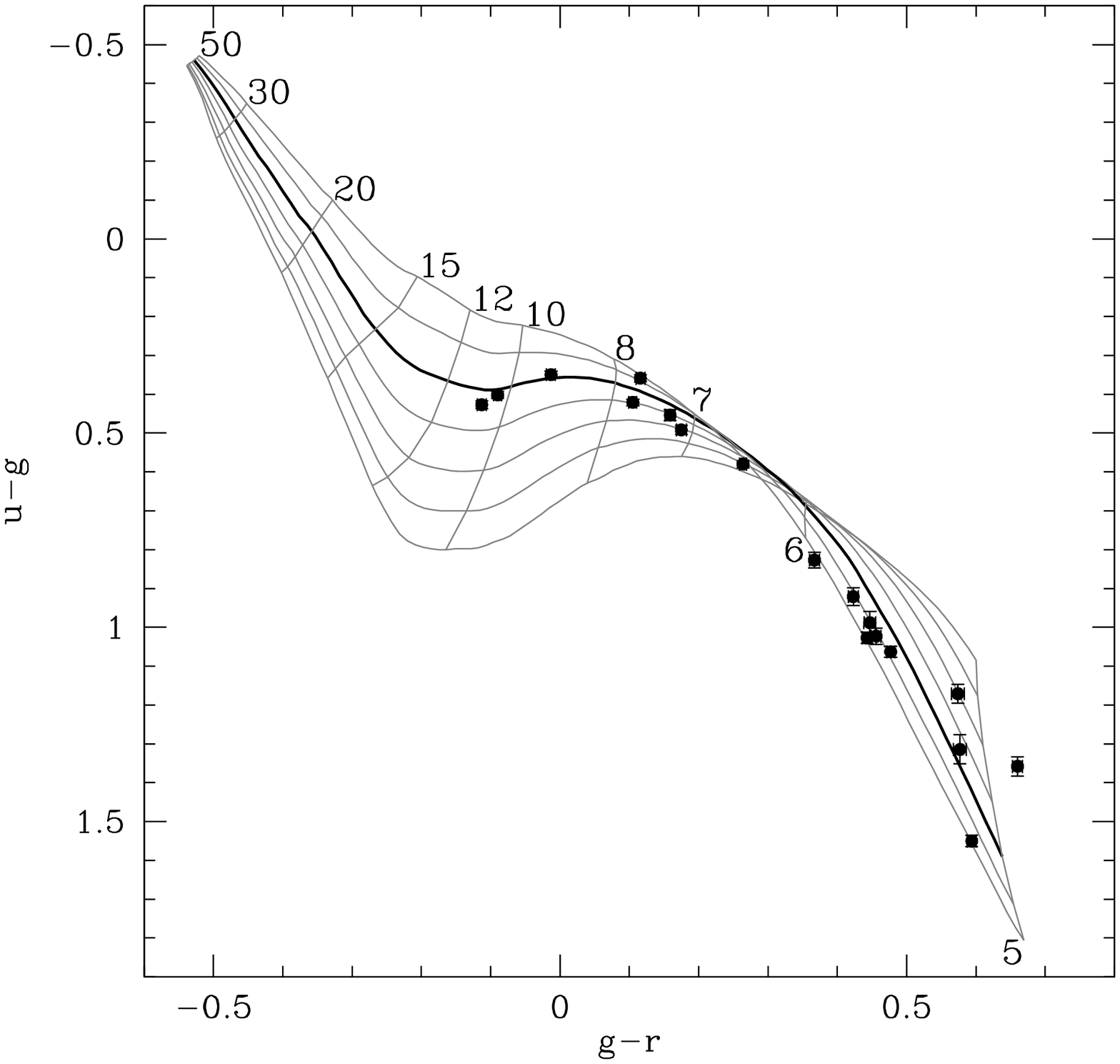}%
\includegraphics[width=\columnwidth]{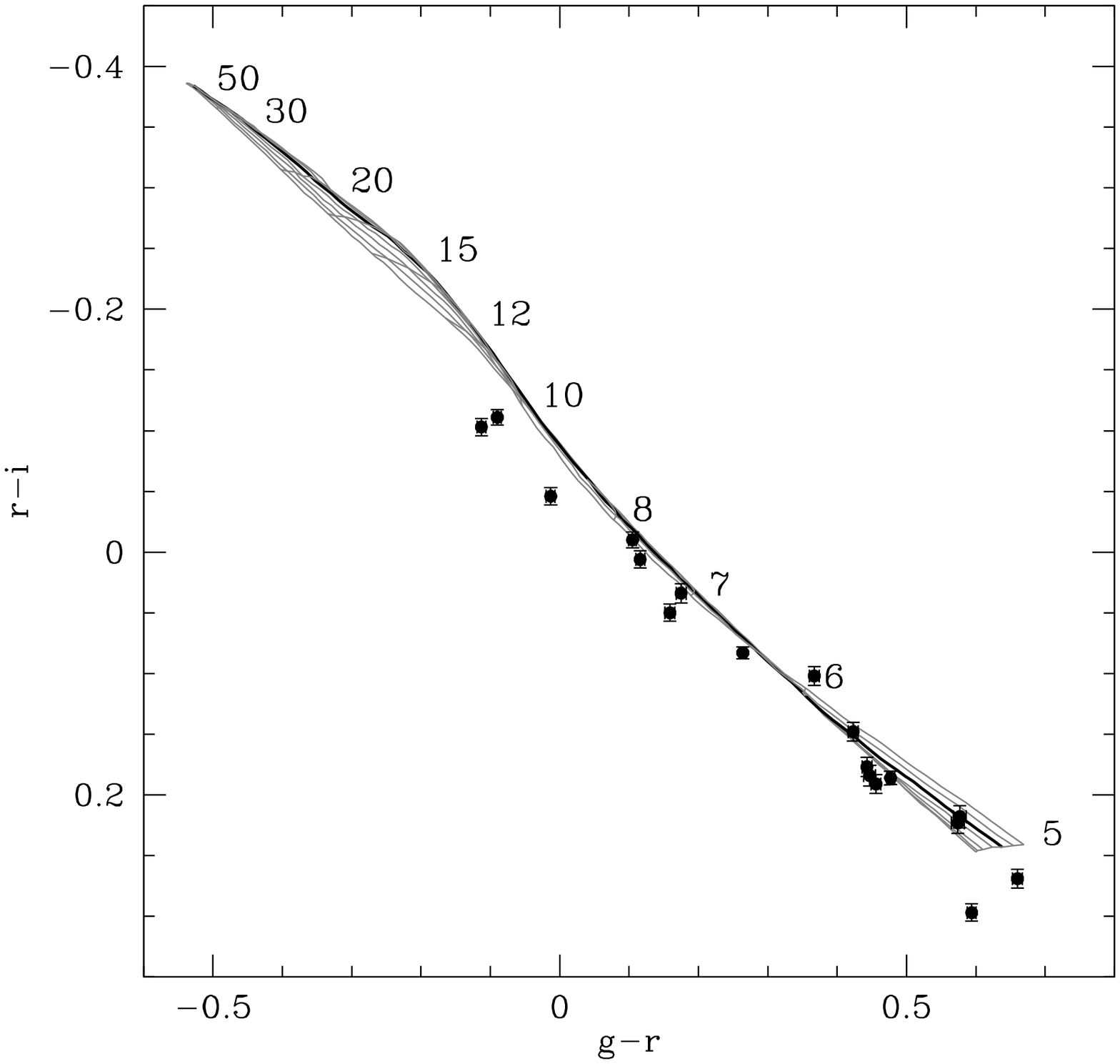}%
\caption{SDSS $u-g$ versus $g-r$ (left panel) and $g-r$ versus $r-i$ 
(right panel) photometry compared to synthetic colours of H-rich white 
dwarfs. The effective temperature is indicated in units of 1000 K and 
$\log{g} = 6.0$ to $9.0$ (in steps of 0.5 from bottom to top).
The thick line indicates $\log{g} = 8$.}
\label{fig_umg_gmr}
\end{figure*}

The hydrogen-rich white dwarfs were analysed for their effective temperature
and surface gravity using a grid of pure-hydrogen LTE plane-parallel
models. The grid of models extends from \teff$\,=4900$ K to 8000 K
(in steps of 100 K), from 8200 to 10000 K (in steps of 200 K) and from 10500 to
20000 K (in steps of 500 K) at
$\log{g} = 5.5$ to 9.5 (in steps of 0.25 dex). The models are described in
\citet{kaw2006} and \citet{kaw2007} with some upgrades described in
\citet{kaw2012}. The observed Balmer line profiles were
fitted with model spectra using $\chi^2$ minimization techniques. The quoted
uncertainties are only statistical ($1\sigma$) and do not take into account
possible systematic effects in model calculations, data acquisition or
reduction procedures.

We used the evolutionary mass-radius relations of \citet{ben1999} to convert the
measured effective temperature and surface gravity measurements into white dwarf ages and masses. For
hydrogen-rich white dwarfs we used the models with $M_H/M_* = 10^{-4}$ and a
metallicity of Z=0. For helium rich white dwarfs, we used the models of
\citet{ben1999} without a hydrogen envelope.
The evolutionary tracks of \citet{ser2002} were employed to interpret the parameters of the
extremely low-mass white dwarf NLTT~11748.

The longitudinal magnetic field was determined by first fitting the Balmer line
profiles with a model profile. We then used this best-fitting model profile to 
calculate the $V/I$ spectra at various magnetic field strengths ($B_{\rm l}$ in Gauss):
\begin{equation}
V/I = \frac{B_{\rm l}\ 4.67\times10^{-13} \lambda^2}{F} \frac{dF}{d\lambda},
\end{equation}
in first order, where $\lambda$ is the wavelength in \AA\ and $F$ is the spectral flux. This grid of
synthetic circular polarization spectra are then fitted to the observed circular 
polarization spectra using $\chi^2$ minimization techniques. 

The procedure applied to our flux calibration standards EG~21 (WD~0310$-$688) and EG~274 (WD~1620$-$391) provided field measurements of 
$B_{\rm l}=-8.1\pm6.2$ and $0.5\pm3.5$ kG, respectively, using the H$\beta$, $\gamma$, and $\delta$ lines.
The errors are comparable to other measurements from this programme owing to the relatively short exposure times employed when observing
standards.
The present results confirm the null results obtained by \cite{kaw2007} who measured $B_{\rm l}=-6.1\pm2.2$ (EG~21) and $-3.0\pm2.6$ kG (EG~274) 
using the H$\alpha$, $\beta$, and $\gamma$ lines.

\subsection{Sample properties}

\subsubsection{\teff\ and $\log{g}$ measurements}

\begin{figure}
\centering
\includegraphics[width=1.0\columnwidth]{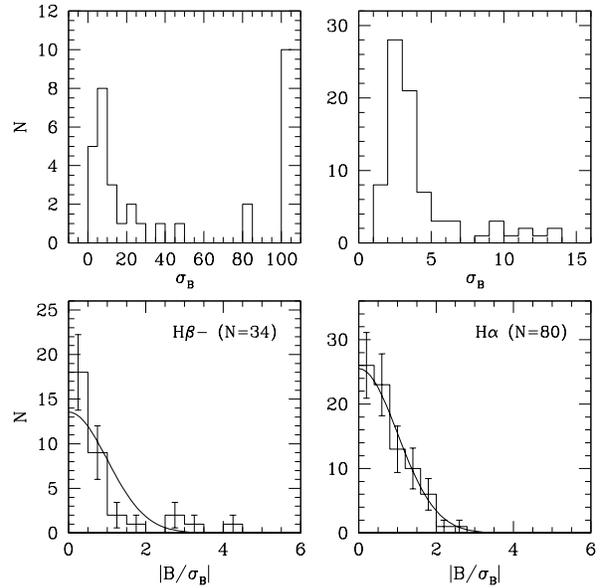}
\caption{Error statistics of the magnetic field measurements: (upper panels)
error distributions in the longitudinal field strength when measuring the upper
Balmer lines (left panels) or H$\alpha$ (right panels), and (lower panels) the corresponding distributions of the measurement
significance ($B/\sigma_{\rm B}$). We obtained 34 individual measurements of the upper Balmer lines and 80 individual
measurements of H$\alpha$ for 20 and 39 stars, respectively.}
\label{fig_stat_error}
\end{figure}

Table~\ref{tbl_atm} lists the measured atmospheric parameters of the sample 
stars. Using Balmer lines alone, the errors in surface gravity measurements 
increase considerably at low temperatures \teff\,$\la 6500$ K. 
Figure~\ref{fig_tg} (left panels) shows the gravity versus temperature 
measurements based on fitting of the Balmer lines. The data set, which 
excludes the high-field DAP white dwarfs NLTT~12758 and 13015, comprises 83 
individual measurements obtained using different instruments described in 
Section 2. The surface gravity distribution, excluding NLTT~11748, is 
consistent with a sample average $<\log{g}>=7.92$ and an intrinsic width 
$\sigma=0.36$ dex ($\chi^2=1$) corresponding to an average mass of 
$0.54_{-0.15}^{+0.23}\, M_\odot$. The average mass is lower than determined for 
younger white dwarf samples such as the SDSS white dwarf sample for which mean 
masses of $0.593\ M_\odot$ \citep{kep2007} and $0.613\ M_\odot$ \citep{tre2011}
were determined. \citet{gia2011} determined a mean mass of $0.661\ M_\odot$ from
their spectroscopic analysis of white dwarfs selected from the McCook \& Sion
catalogue \citep{mcc1999}. Recently, \citet{gia2012} reanalysed the local sample 
of white dwarfs and determined a mean mass of $0.650\ M_\odot$.

Figure~\ref{fig_umg_gmr} shows photometric data of a sub-sample of 18 of these objects and synthetic colours computed
using our model atmosphere grid. The observed colours clearly show the effect of extended Ly$\alpha$ absorption
described by \citet{kow2006}.

\begin{table*}
\centering
\begin{minipage}{\textwidth}
\caption{Atmospheric and other properties. \label{tbl_atm}}
\renewcommand{\footnoterule}{\vspace*{-15pt}}
\centering
\begin{tabular}{llccccccc}
\hline\hline
WD & NLTT & \multicolumn{2}{c}{Adopted parameters \footnote{Based on the analysis of Balmer line profiles except for the high-field DAP white dwarfs NLTT~12758 and 13015 (see text).}} & & & & \\
\cline{3-4} \\
   &      & \teff & $\log{g}$       & Mass  & Age      & $M_V$ & $d$  \\
   &      & ( K )         & (c.g.s)           & $(M_\odot)$ & (Gyr) & (mag) & (pc) \\
\hline
0001$-$039                                      & 00082 & $6910\pm 40$ & $7.67\pm0.08$  & $0.43\pm0.03$   & 1.5 & $13.15\pm0.09$ & $53\pm2$ \\
0005$-$148 \footnote{Possible DAP white dwarf.} & 00347 & $6400\pm180$ & $(8.00\pm0.24)$& $ 0.59\pm0.15 $ & 2.0 & $13.95\pm0.42$ & $42\pm8$ \\
0015$-$055 \footnote{DAZ white dwarf.}          & 00888 & $5680\pm190$ & $(8.00\pm0.24)$& $ 0.59\pm0.15 $ & 2.9 & $14.49\pm0.47$ & $43\pm9$ \\
0038$-$084 \footnote{DAP white dwarf.}          & 02219 & $6000\pm180$ & $(8.00\pm0.24)$& $ 0.59\pm0.15 $ & 2.4 & $14.23\pm0.44$ & $41\pm8$ \\
0049$-$308                                      & 02886 & $6440\pm180$ & $(8.00\pm0.24)$& $ 0.59\pm0.15 $ & 2.0 & $13.92\pm0.42$ & $55\pm6$ \\
0053$-$117                                      & 03080 & $6840\pm 40$ & $7.72\pm0.09$  & $0.44\pm0.04$   & 1.3 & $13.28\pm0.13$ & $25\pm2$ \\
0100$-$036                                      & 03471 & $8720\pm 60$ & $7.96\pm0.08$  & $0.57\pm0.05$   & 0.84& $12.66\pm0.12$ & $61\pm3$ \\
0120$-$024 \footnote{In a common proper-motion binary with a main-sequence star.} & 04615 & $5840\pm210$ & $(8.00\pm0.24)$& $ 0.59\pm0.15 $ & 2.6 & $14.35\pm0.47$ & $42\pm9$ \\
0136$-$042                                      & 05503 & $9870\pm 50$ & $8.12\pm0.04$  & $0.67\pm0.02$   &0.78 & $12.43\pm0.06$ & $60\pm2$ \\
0136$-$340                                      & 05543 & $6620\pm160$ & $(8.00\pm0.24)$& $ 0.59\pm0.15 $ & 1.8 & $13.80\pm0.40$ & $47\pm9$ \\
0142$-$304                                      & 05814 & $10490\pm120$& $7.97\pm0.08$  & $0.58\pm0.05$   &0.53 & $11.98\pm0.13$ & $79\pm5$ \\
0145$-$274 \vspace*{-1.5pt}                     & 06004 & $6620\pm230$ & $(8.00\pm0.24)$& $ 0.59\pm0.15 $ & 1.8 & $13.80\pm0.45$ & $32\pm7$ \\
0151$-$308 \large{$^c$}                         & 06390 & $6050\pm180$ & $(8.00\pm0.24)$& $ 0.59\pm0.15 $ & 2.4 & $14.19\pm0.44$ & $42\pm8$ \\
0155$-$070                                      & 06559 & $10600\pm150$& $7.83\pm0.09$  & $0.51\pm0.05$   &0.42 & $11.75\pm0.16$ & $84\pm6$ \\
\vspace*{-1.5pt} 0159$-$270                     & 06794 & $7480\pm 50$ & $8.30\pm0.08$  & $0.78\pm0.05$   & 2.3 & $13.77\pm0.10$ & $49\pm2$ \\
\vspace*{-1.5pt} 0204$-$306 \large{$^e$}        & 07051 & $5640\pm200$ & $(8.00\pm0.24)$& $ 0.58\pm0.15 $ & 2.9 & $14.52\pm0.48$ & $30\pm7$ \\
0214$-$071 \large{$^c$}                         & 07547 & $5410\pm200$ & $(8.00\pm0.24)$& $ 0.58\pm0.15 $ & 3.4 & $14.73\pm0.50$ & $44\pm10$\\
0233$-$230                                      & 08432 & $6440\pm200$ & $(8.00\pm0.24)$& $ 0.59\pm0.15$  & 2.0 & $13.92\pm0.43$ & $45\pm9$ \\
0304$-$074                                      & 09940 & $5660\pm270$ & $(8.00\pm0.24)$& $ 0.58\pm0.15 $ & 2.9 & $15.50\pm0.46$ & $37\pm9$ \\
0315$-$293 \footnote{Magnetic DAZ (DAZP) white dwarf \citep{kaw2011}.} & 10480 & $5340\pm190$ & ($8.00\pm0.24$) & $ 0.58\pm0.15 $ & 3.7 & $14.79\pm0.49$    & $33\pm8$  & \\
0322$-$175 \vspace*{-1.5pt}                     & 10884 & $6884\pm 42$ & $7.66\pm0.09$  & $0.42\pm0.03$   & 1.6 & $13.15\pm0.10$ & $64\pm3$ \\
0334$-$224 \large{$^c$}                         & 11393 & $5930\pm170$ & $(8.00\pm0.24)$& $ 0.59\pm0.15 $ & 2.5 & $14.28\pm0.43$ & $36\pm7$ \\
0342$+$176 \vspace*{-1.5pt}                     & 11748 & $8412\pm 42$ & $6.38\pm0.10$  & $0.17\pm0.01$   & 4.5 & $10.11\pm0.07$ & $204\pm7$ \\ 
0410$-$114 \large{$^d$}                         & 12758 & $7440\pm150$ & $(8.00\pm0.24)$& $ 0.59\pm0.15 $ & 1.3 & $13.34\pm0.38$ & $ 26\pm5 $ \\
0411$-$081 \vspace*{-1.5pt}                     & 12796 & $8830\pm40$  & $8.04\pm0.06$  & $0.62\pm0.04$   & 0.92& $12.73\pm0.09$ & $51\pm2$ \\
0416$-$096 \large{$^d$}                         & 13015 & $5745\pm405$ & $(8.00\pm0.24)$& $ 0.59\pm0.15 $ & 2.8 & $ 14.43\pm0.66 $ & $ 40\pm12 $ \\
0429$-$194                                      & 13471 & $5780\pm210$ & $(8.00\pm0.24)$& $ 0.59\pm0.15 $ & 2.7 & $14.40\pm0.48$ & $34\pm7$ \\
0431$-$279                                      & 13532 & $5300\pm220$ & $(8.00\pm0.24)$& $ 0.58\pm0.15 $ & 3.8 & $14.83\pm0.53$ & $24\pm6$  \\
0438$-$154                                      & 13755 & $6000\pm160$ & $(8.00\pm0.24)$& $ 0.59\pm0.15 $ & 2.4 & $14.23\pm0.42$ & $33\pm8$ \\
0503$-$175                                      & 14491 & $6140\pm180$ & $(8.00\pm0.24)$& $ 0.59\pm0.15 $ & 2.3 & $14.13\pm0.43$ & $41\pm8$ \\
0506$-$154                                      & 14558 & $5360\pm210$ & $(8.00\pm0.24)$& $ 0.58\pm0.15 $ & 3.6 & $14.78\pm0.51$ & $39\pm9$ \\
0556$-$042                                      & 15882 & $6100\pm190$ & $(8.00\pm0.24)$& $ 0.59\pm0.15 $ & 2.3 & $14.16\pm0.44$ & $40\pm8$ \\
0559$+$158                                      & 15957 & $6870\pm 70$ & $8.05\pm0.13$  & $0.62\pm0.08$ & 1.8  & $13.72\pm0.19$ & $41\pm4$ \\
0707$-$320                                      & 17486 & $9903\pm 52$ & $7.98\pm0.05$  & $0.59\pm0.03$ & 0.62 & $12.20\pm0.08$ & $48\pm2$ \\
0715$+$125A                                     & 17662A& $7560\pm120$ & $8.12\pm0.21$  & $0.67\pm0.14$ & 1.6  & $13.45\pm0.30$ & $33\pm4$ \\
0715$+$125B                                     & 17662B& $6025\pm75$  & $8.25\pm0.25$  & $0.75\pm0.16$ & 3.6  & $14.58\pm0.30$& ...  \\
0724$+$146                                      & 17874 & $5780\pm240$ & $(8.00\pm0.24)$& $ 0.59\pm0.15 $ & 2.7 & $14.40\pm0.50$ & $31\pm7$ \\
0925$+$189                                      & 21844 & $7487\pm 30$ & $8.22\pm0.04$  & $0.73\pm0.03$ & 2.0 & $13.64\pm0.05$ & $40\pm1$ \\
0927$-$173A \footnote{DA+DC binary: (DC) \teff$=6150\pm150$ K, $\log{g}=8.11\pm0.12$, $M=0.63\pm0.08\,M_\odot$.} & 21913 & $9550\pm115$ & $8.58\pm0.07$ & $ 0.96\pm0.04 $   & 1.8  & $13.31\pm0.10$ & $43\pm2$ \\
1016$-$044                                      & 23966 & $7790\pm 40$ & $7.88\pm0.07$  & $0.52\pm0.04$ & 0.99 & $12.99\pm0.10$ & $67\pm4$ \\
1025$+$197 \vspace*{-1.5pt}                     & 24406 & $6959\pm 73$ & $8.13\pm0.12$  & $0.67\pm0.08$ & 2.0 & $13.79\pm0.17$ & $51\pm4$ \\
1054$-$226 \large{$^c$}                         & 25792 & $7921\pm 33$ & $7.88\pm0.06$  & $0.52\pm0.03$ &0.95 & $12.92\pm0.09$ & $41\pm2$ \\
1143$-$013                                      & 28493 & $6500\pm220$ & $(8.00\pm0.24)$& $ 0.59\pm0.15 $ & 1.9 & $13.88\pm0.44$ & $32\pm7$ \\
\vspace*{-1.5pt} 1150$-$066                     & 28730 & $8595\pm 28$ & $8.44\pm0.04$  & $0.87\pm0.03$ & 2.0 & $13.47\pm0.05$ & $41\pm1$ \\
1209$-$060 \large{$^e$}                         & 29967 & $6400\pm170$ & $(8.00\pm0.24)$& $ 0.59\pm0.15 $ & 2.0 & $13.95\pm0.41$ & $41\pm8$ \\
1237$-$230                                      & 31473 & $5680\pm150$ & $(8.00\pm0.24)$& $ 0.59\pm0.15 $ & 2.9 & $14.49\pm0.43$ & $26\pm5$ \\
1238$+$183                                      & 31483 & $5500\pm240$ & $(8.00\pm0.24)$& $ 0.58\pm0.15 $ & 3.2 & $14.65\pm0.53$ & $39\pm10$ \\
1302$-$052                                      & 32785 & $5340\pm140$ & $(8.00\pm0.24)$& $ 0.58\pm0.15 $ & 3.7 & $14.79\pm0.44$ & $30\pm6$ \\
1314$-$153                                      & 33503 & $14350\pm120$& $7.81\pm0.04$  & $0.51\pm0.02$ & 0.17& $11.04\pm0.06$ & $58\pm2$ \\
1316$-$215                                      & 33669 & $5690\pm150$ & $(8.00\pm0.24)$& $ 0.59\pm0.15 $ & 2.8 & $14.48\pm0.43$ & $28\pm6$ \\
1350$-$274                                      & 35570 & $7570\pm 50$ & $7.95\pm0.08$  & $0.56\pm0.05$ & 1.2 & $13.20\pm0.12$ & $56\pm3$ \\
1444$-$303                                      & 38356 & $6950\pm 30$ & $7.78\pm0.05$  & $0.47\pm0.03$ & 1.2 & $13.31\pm0.08$ & $63\pm2$ \\
2053$-$274                                      & 50161 & $8880\pm 40$ & $8.09\pm0.05$  & $0.65\pm0.03$ &0.98 & $12.78\pm0.07$ & $74\pm3$ \\
\hline
\end{tabular}
\end{minipage}
\end{table*}

\begin{table*}
\centering
\begin{minipage}{\textwidth}
\contcaption{}
\renewcommand{\footnoterule}{\vspace*{-15pt}}
\centering
\begin{tabular}{llccccccc}
\hline\hline
WD & NLTT & \multicolumn{2}{c}{Adopted parameters \footnote{Based on the analysis of Balmer line profiles.}}  & & & \\
\cline{3-4} \\
   &      & \teff & $\log{g}$       &    Mass & Age  & $M_V$ & $d$  \\
   &      & ( K )         & (c.g.s)           & $(M_\odot)$ & (Gyr)& (mag) & (pc) \\
\hline
2123$-$229 & 51252     & $13605\pm127$& $7.95\pm0.03$   & $0.58\pm0.02$ & 0.25& $11.35\pm0.05$ & $60\pm2$ \\
2308$-$300A& 56045     & $7520\pm 60$ & $7.77\pm0.11$   & $0.47\pm0.06$ &0.94 & $12.98\pm0.17$ & $57\pm5$ \\
2312$-$149 & 56257     & $8770\pm 50$ & $8.53\pm0.06$   & $0.93\pm0.04$ & 2.1 & $13.56\pm0.07$ & $26\pm1$ \\
2317$-$027 & 56493     & $5300\pm220$ & $(8.00\pm0.24)$ & $ 0.58\pm0.15 $ & 3.8 & $ 14.83\pm0.53 $ & $34\pm8$ \\
2322$+$137 & 56805     & $5300\pm140$ & $(8.00\pm0.24)$ & $ 0.58\pm0.15 $ & 3.8  & $ 14.83\pm0.44 $& $17\pm4$ \\
1145$-$451 & ...       & $6100\pm150$ & $(8.00\pm0.24)$ & $ 0.59\pm0.15 $ & 2.3  & $14.16\pm0.41$ & $20\pm4$ \\
\hline
\end{tabular}
\end{minipage}
\end{table*}

Several measurements based on Balmer lines, particularly those excluding H$\alpha$, are unsatisfactory below 
$\sim 6500$ K. In such cases where the error on the surface gravity exceeded 
0.5 dex, we adopted the average and dispersion of gravity measurements for 
stars with temperatures above $\sim 6500$ K, i.e,, $<\log{g}>=8.0$ and 
$\sigma_g=0.24$, and we fitted the Balmer lines within this constraint to determine the temperature. Next, we compared the adopted effective temperatures to 
measurements based on SDSS $ugriz$ photometry and on the $V-J$ and $R-J$ 
indices: Figure ~\ref{fig_tg} (right panel) show the measurements obtained 
using the various methods. The results based on Balmer lines and the $V-J$ 
index differ by 16~K only with a dispersion of 412 K. Those based on the $R-J$ 
index differ from the Balmer line results by 72~K with a dispersion of 552~K. 
The temperatures measured using the SDSS $ugriz$ data differ by $-117$~K on 
average but with a lower dispersion of 189~K. The latter also suggest that 
the effective temperature of cool white dwarfs ($\la 5500$ K) obtained by 
fitting Balmer lines assuming $\log{g}=8$ may be overestimated by a few 100~K.

Table~\ref{tbl_atm} also lists the absolute magnitude, age, mass, and distance estimates. The distances
were obtained by calculating the distance modulus $m-M$ from the absolute
and apparent magnitudes (Appendix A).
The DA white dwarfs NLTT~3080, NLTT~13532,
NLTT~56805, and PM~J11480$-$4523 probably reside within 25~pc from the Sun. 

\subsubsection{Magnetic field measurements}

\begin{table}
\centering
\begin{minipage}{\columnwidth}
\caption{Magnetic field measurements - detections. \label{tbl_mag_meas}}
\renewcommand{\footnoterule}{\vspace*{-15pt}}
\centering
\begin{tabular}{lccc}
\hline\hline
 NLTT & $B_{\rm l}$ & $B_{\rm S}$ & lines \\
      & (kG)  & (kG)        &       \\
\hline
  347 & $-4.2\pm2.8$ &     & H$\alpha$       \\
      & $-5.2\pm2.8$ &     & H$\alpha$       \\
      & \boldmath{$-4.6\pm1.9$} & ... & H$\alpha$  \\  
 2219 & $-112.8\pm38.8$ &   & H$\beta$,H$\gamma$,H$\delta$ \\
      &  $-79.8\pm26.6$ &   & H$\beta$,H$\gamma$,H$\delta$  \\
      & \boldmath{$-97.0\pm21.6$} & $\sim 300.$  & H$\beta$,H$\gamma$,H$\delta$ \\
10480\footnote{From \citet{kaw2011}.} & $-212.0\pm50.0$ & $519.\pm4.$ & Ca~H\&K \\
12758 & ... & $(1.7\pm0.2)\times10^3$ & H$\alpha$ \\
13015 & ... & $(6.0-7.5)\times10^3$ & H$\alpha$ \\
\hline
\end{tabular}\\
\end{minipage}
\end{table}

Probable or definite non-zero field measurements are reported in Table~\ref{tbl_mag_meas} where
bold-faced entries are weighted averages, 
and Table~\ref{tbl_nonmag_meas} lists average field measurements that are statistically consistent
with the null hypothesis. The uncertainties vary and are
lower when measuring strong, narrow lines. For example, the error obtained in measuring
the Ca~H\&K lines in NLTT~7547 is a factor of four lower than obtained in measuring the shallow
Balmer lines. 

Figure~\ref{fig_stat_error} shows the error distributions obtained using the
H$\alpha$ or the upper-Balmer set-ups. The sample depicted in the diagram excludes
the strong-field objects NLTT~12758 and NLTT~13015. The errors are typically of the order of
2-3 kG when measuring H$\alpha$, and the errors increase to 5-10 kG when
measuring the upper Balmer lines. The field distribution obtained using the upper Balmer lines
also shows a systematic offset of 4~kG. The significance of the measurements,
$B/\sigma_{\rm B}$, closely follows a normalized Gaussian distribution when
measuring H$\alpha$, but the distribution obtained when measuring the less significant
upper Balmer lines departs somewhat from expectations indicating that the errors
were overestimated by $\la 50$\%. Two outliers stood out, NLTT~2219 in the upper-Balmer distribution
and NLTT~347 in the H$\alpha$ distribution.

\begin{table*}
\centering
\begin{minipage}{\textwidth}
\caption{Magnetic field measurements - non-detections ($N=54$). \label{tbl_nonmag_meas}}
\renewcommand{\footnoterule}{\vspace*{-15pt}}
\centering
\begin{tabular}{rccrcc}
\hline\hline
 NLTT & $B_{\rm l}$ & lines & NLTT & $B_{\rm l}$ & lines \\
      & (kG)  &       &      & (kG)  &  \\
\hline
   82 & $0.6\pm1.6$ & H$\alpha$                      & 15957 & $-0.8\pm1.6$ & H$\alpha$ \\
  888 & $-3.9\pm39.8$ & H$\beta$,H$\gamma$,H$\delta$ & 17486 & $0.4\pm3.1$ & H$\alpha$ \\
 2886 & $-1.5\pm1.9$ & H$\alpha$                     & 17662A & $-3.2\pm2.0$ & H$\alpha$ \\
 3080 & $-0.7\pm1.2$ & H$\alpha$                     & 17662B & $-3.5\pm4.7$ & H$\alpha$ \\
 3471 & $0.2\pm2.1$  & H$\alpha$                     & 17874 & $4.6\pm2.3$ & H$\alpha$ \\
 4615 & $-0.3\pm2.4$ & H$\alpha$                     & 21844 & $0.7\pm4.8$ & H$\beta$,H$\gamma$,H$\delta$ \\
 5503 & $2.0\pm2.5$  & H$\alpha$                     & 21913 & $-1.6\pm2.2$ & H$\alpha$ \\
 5543 & $0.0\pm5.2$  & H$\beta$,H$\gamma$,H$\delta$  & 23966 & $-3.4\pm2.2$ & H$\alpha$ \\
 5814 & $-0.3\pm5.0$ & H$\alpha$                     & 24406 & $3.0\pm2.4$ & H$\alpha$ \\
 6004 & $6.8\pm9.4$ & H$\beta$,H$\gamma$,H$\delta$   & 25792 & $3.9\pm3.0$ & H$\beta$,H$\gamma$,H$\delta$ \\
 6390 & $-3.9\pm10.8$ & H$\beta$,H$\gamma$,H$\delta$ & 28493 & $20.6\pm7.4$ & H$\beta$,H$\gamma$,H$\delta$ \\
 6559 & $-3.2\pm2.2$  & H$\alpha$                    & 28730 & $-2.4\pm2.0$ & H$\beta$,H$\gamma$,H$\delta$ \\
 6794 & $-3.2\pm2.0$  & H$\alpha$                    & 29967 & $-3.6\pm6.5$ & H$\beta$,H$\gamma$,H$\delta$ \\
 7051 & $-1.2\pm6.4$ & H$\alpha$                     & 31473 & $1.0\pm1.8$ & H$\alpha$ \\
 7547 & $-23.5\pm177.0$ & H$\beta$,H$\gamma$         & 31483 & $-2.6\pm80.5$ & H$\beta$,H$\gamma$ \\
      & $-32.2\pm40.2$ & Ca~H\&K                   & 32785 & $9.2\pm6.3$ & H$\alpha$ \\
 8432 & $-0.4\pm2.2$ & H$\alpha$                     & 33503 & $1.6\pm1.8$ & H$\alpha$ \\
 9940 &  $4.0\pm4.5$ & H$\alpha$                     & 33669 & $-2.8\pm24.8$ & H$\beta$,H$\gamma$,H$\delta$ \\
10884 & $-1.3\pm12.2$& H$\beta$,H$\gamma$,H$\delta$  & 35570 & $-3.7\pm2.9$ & H$\alpha$ \\
11393 & $7.8\pm15.8$ & H$\beta$,H$\gamma$,H$\delta$  & 38356 & $-4.4\pm4.1$ & H$\beta$,H$\gamma$,H$\delta$ \\
11748 & $-2.6\pm2.7$  & H$\alpha$                     & 50161 & $-4.1\pm5.7$ & H$\alpha$ \\                    
12796 & $-2.3\pm2.2$ & H$\alpha$                     & 51252 & $-0.3\pm2.5$ & H$\alpha$ \\
13471 & $-0.1\pm2.1$ & H$\alpha$                     & 56045 & $2.8\pm3.4$ & H$\alpha$ \\ 
13532 & $-5.2\pm6.3$ & H$\alpha$                     & 56257 & $-2.3\pm1.7$ & H$\alpha$ \\
13755 &  $0.8\pm1.4$ & H$\alpha$                     & 56493 & $-1.9\pm13.4$ & H$\alpha$ \\
14491 &  $0.4\pm1.7$ & H$\alpha$                     & 56805 & $-333.9\pm616.4$ & H$\beta$ \\
14558 &  $7.5\pm108.0$ & H$\beta$                    & PMJ11480$-$4523 & $0.4\pm10.5$ & H$\beta$,H$\gamma$,H$\delta$ \\
15882 &  $2.3\pm1.5$ & H$\alpha$                     &                 &              &                              \\
\hline
\end{tabular}\\
\end{minipage}
\end{table*}

\subsection{New low-field magnetic white dwarfs}

\begin{figure*}
\includegraphics[width=\columnwidth]{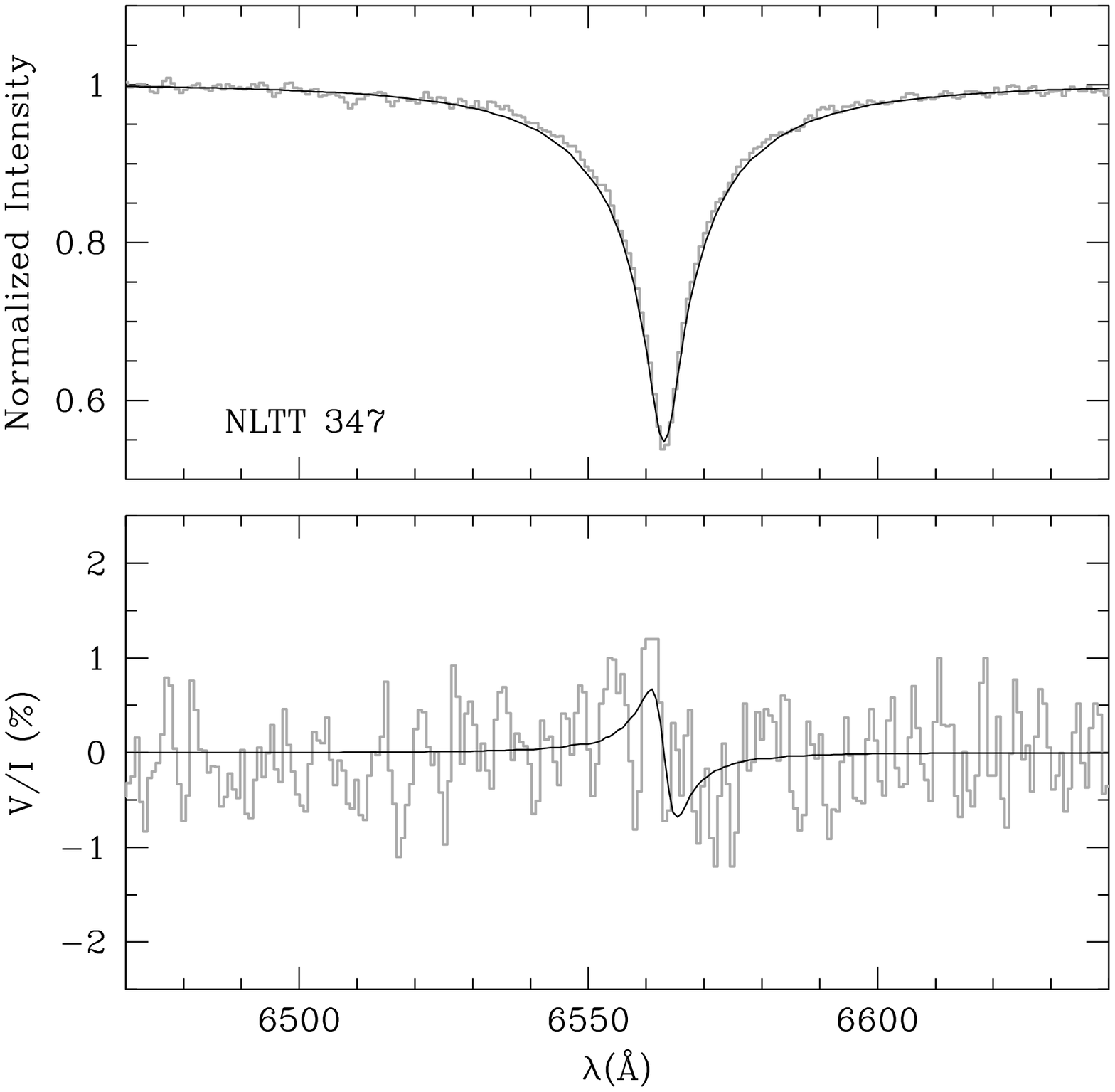}
\includegraphics[width=\columnwidth]{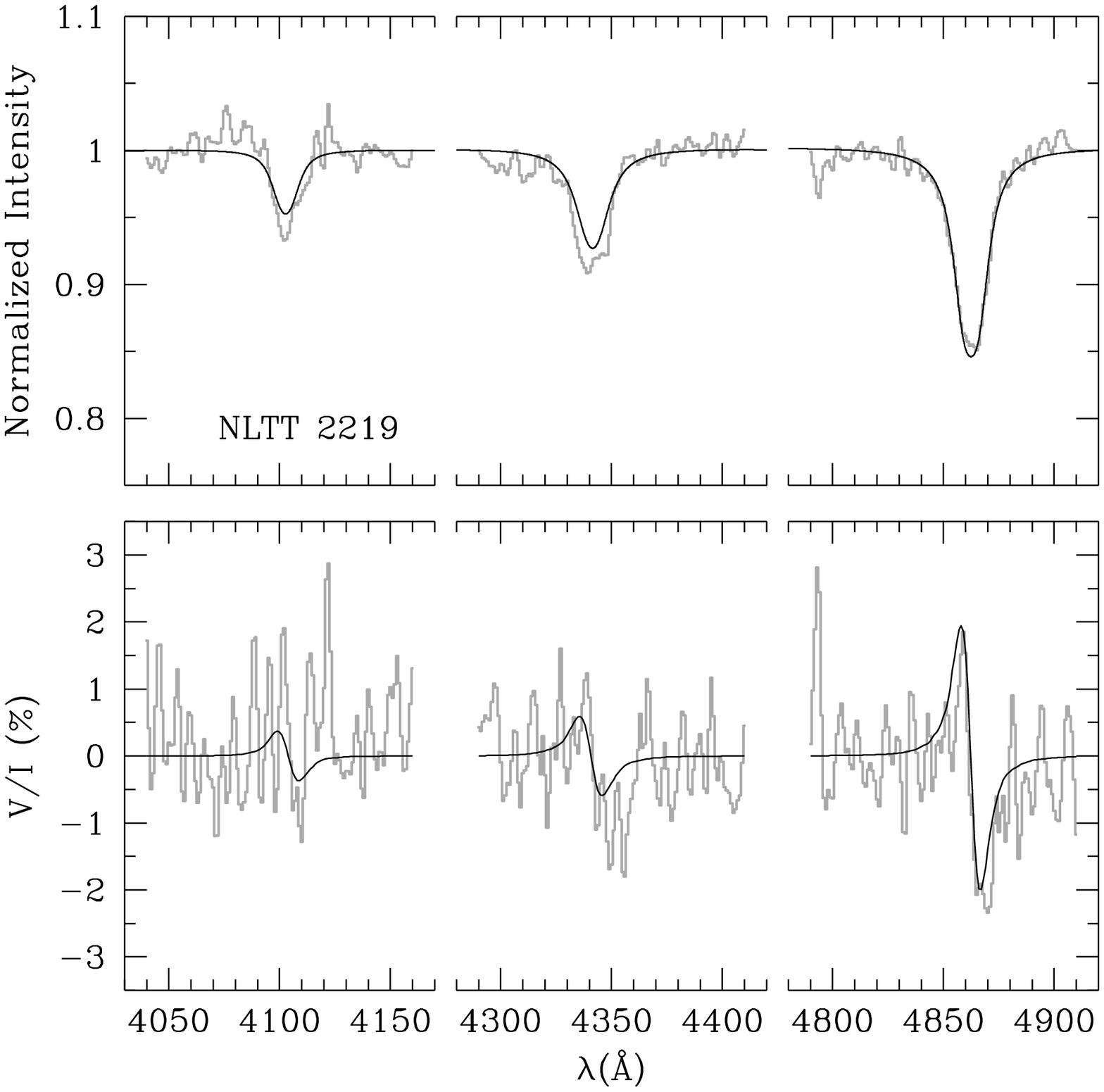}
\caption{Flux (top left) and circular polarization (bottom left) spectra of 
NLTT~347. The flux spectrum (grey) is compared to the best fit line profile 
(black). The circular polarization spectrum (grey) is compared to a model 
polarization spectrum (black) at $B_{\rm l} = -4.6$ kG.
Flux (top right) and circular polarization (bottom right) spectra
of NLTT~2219. The circular polarization spectra are compared to model
polarization spectra at $B_{\rm l} = -97$ kG.
\label{fig_nltt347_2219}}
\end{figure*}

NLTT~347 is most likely a new weak-field magnetic white dwarf. We measured
a longitudinal field of $-4.6\pm1.9$ kG field in the co-added spectrum, 
which is a $2.4\sigma$ detection. Figure~\ref{fig_nltt347_2219} (left) shows
the combined flux and circular polarization spectra of H$\alpha$ compared
to the best fit models. For individual measurements we obtained
a similar field-strength although at a lower significance.

NLTT 2219 is a cool magnetic white dwarf. Figure~\ref{fig_nltt347_2219} (right) shows
the flux and polarization spectra. The flux spectrum shows a
rounded core due to the weak magnetic field, and the circular polarization spectrum
clearly shows the effect of a weak magnetic field.

We reported previously on NLTT~10480 as a new magnetic ($B_{\rm S} = 519\pm4$ kG) DAZ white dwarf \citep{kaw2011}.
The polarization and flux spectra of Ca~H\&K constrained $B_{\rm l} = -212\pm50$ kG
\citep{kaw2012} suggesting a field inclination of $\approx 60^\circ$. This object
is one of the coolest white dwarfs showing heavy element lines.

\subsection{New, variable, high-field white dwarfs}

\begin{figure*}
\includegraphics[width=\columnwidth]{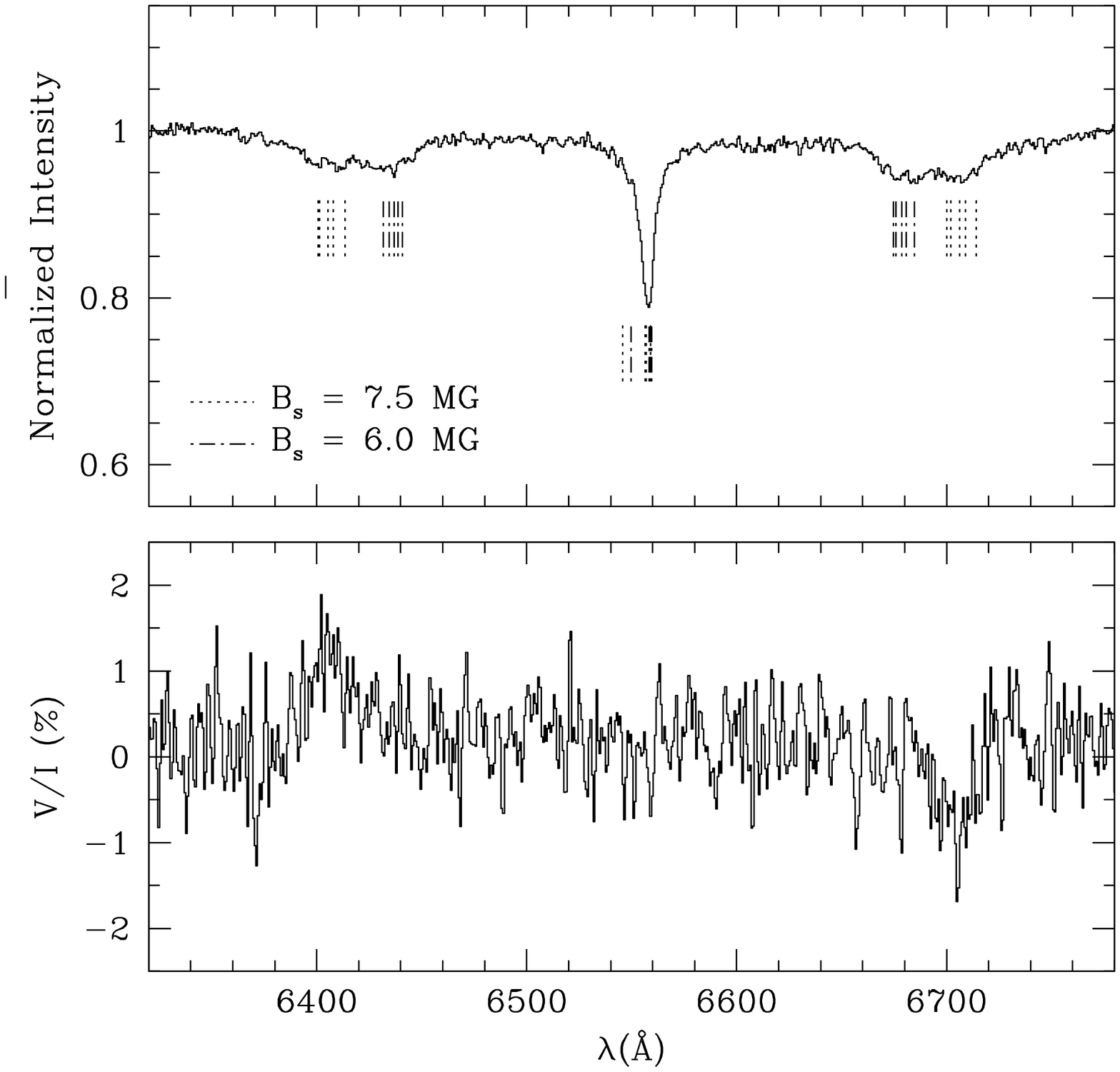}
\includegraphics[width=\columnwidth]{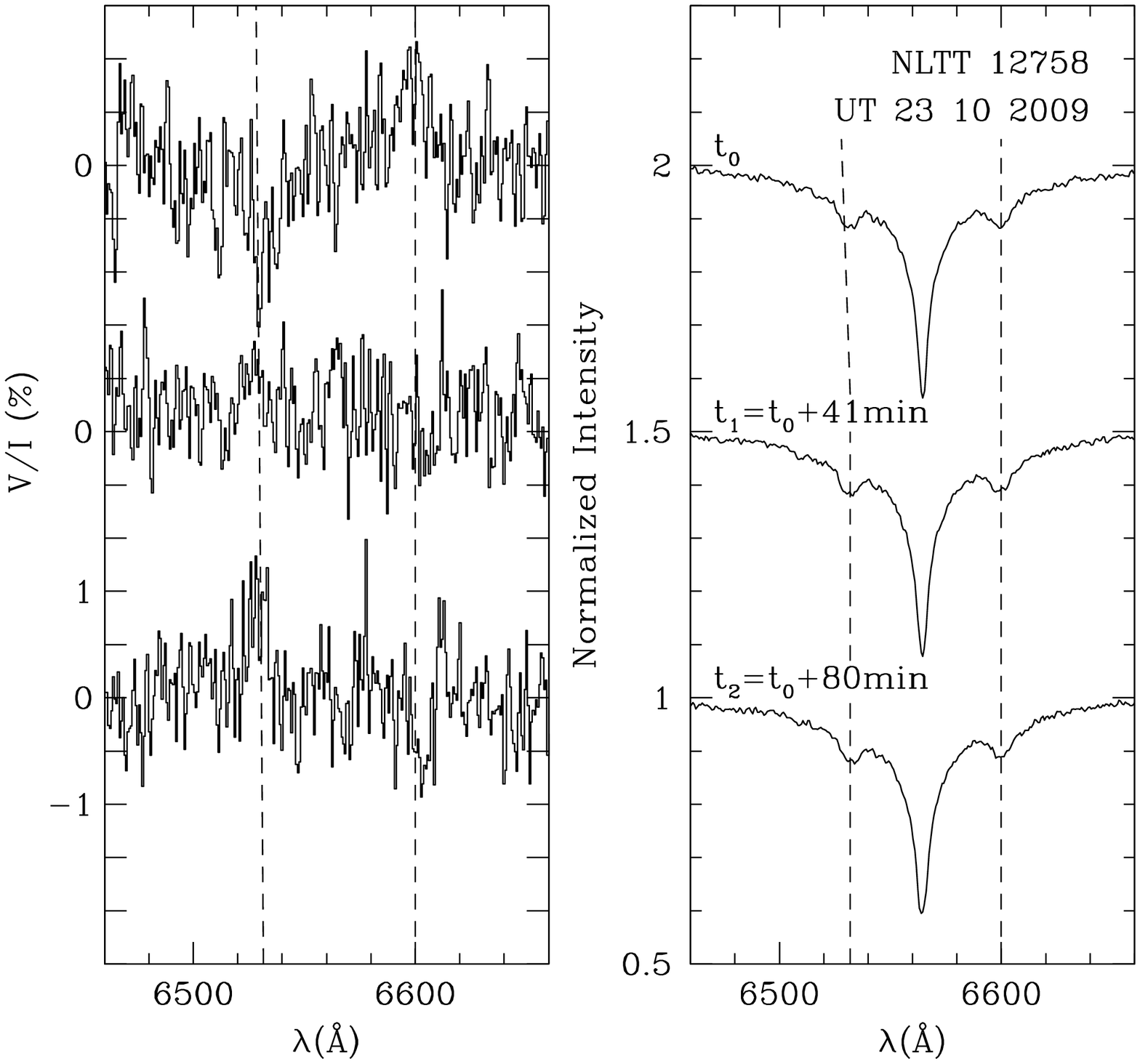}
\caption{(Upper and lower left panels) intensity and circular polarization ($V/I$) spectra of the magnetic DA white dwarf
NLTT~13015. The intensity spectrum shows a field spread of $\sim6-7.5$ MG and relevant line positions from
\citet{kem1974} are marked. Intensity (right panel) and circular
polarization spectrum (middle panel) of the magnetic DA white dwarf NLTT~12758 showing changes in field orientation
within $\sim$80 minutes.
\label{fig_nltt12758_13015}}
\end{figure*}

NLTT~12758 is a new magnetic white dwarf (Fig.~\ref{fig_nltt12758_13015}). We estimated a temperature of
\teff$\,=7440\pm150$ K using the $V-J$ index assuming a gravity $\log{g}=8$. The Zeeman splitting corresponds to a surface
magnetic field of $1.7\pm0.2$ MG. Available photometry ($V = 15.46$, $U-B = -0.71$,
$B-V = +0.31$) from \citet{egg1968} suggests an effective temperature of
$\approx 7700$ K, which is in agreement with our determination.
The structure of the H$\alpha$ profile of NLTT~12758 appears to be similar to
that of the magnetic white dwarf NLTT~48454 (WD~1953$-$011). \citet{max2000}
found that the magnetic field strength of NLTT~48454 varied as shown by variable
Zeeman line splitting in the H$\alpha$ core. They proposed a two-component
model for the magnetic field structure of the white dwarf, whereby the star has
a weak dipolar magnetic field of $\sim 70$~kG with a much stronger spot-like
field of $\sim 500$~kG.
Moreover, as in the case of NLTT~48454, the $\pi$ component is too strong for the
$\sigma$ components suggesting the simultaneous presence of high- and low-
field structures, or that NLTT~12758 is in fact a double degenerate star. Variations in the radial velocity measured using the $\pi$ component
also suggest the presence of a close companion (see Section 3.4). A few resolved magnetic plus non-magnetic double degenerates are known
\citep[see, e.g.,][]{gir2010,dob2012}, while some are, to date, unresolved 
\citep[e.g., EUVE~J1439+75.0][]{ven1999,sch2003}.
NLTT~12758 may be the first known close double degenerate with a magnetic component.

NLTT~13015 is also a new magnetic DA white dwarf (Fig.~\ref{fig_nltt12758_13015}) with a temperature of
$5745\pm405$ (assuming $\log{g}=8$) determined using the $V-J$ and $R-J$ indices. We measured a spread in surface magnetic
field strength ranging from $6.0$ to $7.5$ MG. The field is possibly time variable and quadratic Zeeman effects are apparent in
$\pi$ and $\sigma$ components.

\subsection{Radial velocities and kinematics}

The H$\alpha$ intensity spectra were well suited for a survey of radial 
velocity ($v_r$) variability, and all FORS spectra were used in a detailed kinematical study of this 
sample of high proper-motion white dwarfs. We measured $v_r$ by fitting a 
Gaussian function to the narrow H$\alpha$ or H$\beta$ line cores, or, when 
available, the Ca~H\&K lines. The errors were estimated by varying the flux 
continuum placement. The average and dispersion of the H$\alpha$ velocity 
measurements are $\bar{v}=59.3$ and $\sigma_v=35.3$\kms, while for the 
H$\beta$/Ca~H\&K lines they are $\bar{v}=50.1$ and $\sigma_v=43.6$\kms. 
Individual H$\alpha$ velocity  measurements are listed in Appendix B. Radial 
velocities for NLTT~11748 are listed in \citet{kaw2010}.

We tested the null hypothesis, i.e., a constant velocity, by plotting the 
distribution of velocity measurements relative to the mean for each star, 
$(v_i-\bar{v})/\sigma_v$, where $\bar{v}$ is the weighted average of the 
individual measurements $v_i$. The distribution is then compared to a 
normalized Gaussian function and outliers are identified. First, we noted that 
the distribution was too broad and that systematic errors, i.e.,
instrumental artefacts, probably dominated statistical errors when measuring 
H$\alpha$ velocities and to a lesser extent when measuring the less-accurate 
upper Balmer lines. Therefore, we added a 5\kms\ systematic error in 
quadrature, $\sigma_v^2 + (5\,{\rm km\,s}^{-1})^2$, and recovered a normal 
distribution. Similarly, a normal distribution is recovered for the 
H$\beta$/Ca~H\&K measurements by adding 10\kms\ in quadrature. Apart from the close double 
degenerate NLTT~11748 \citep[see][]{kaw2010} and the peculiar DAP NLTT~12758 mentioned above (Section 3.3), two other objects stood out 
($\chi^2 > 3$) in this survey as possible close binaries, NLTT~6559 and 
NLTT~31473. However, many sequences cover only a short time interval 
($\la 1$ hr) and other likely candidates may emerge (e.g., NLTT~21813, see below).

A comparison with published velocity measurements demonstrates the reliability of the 
H$\alpha$ velocity scale. \citet{zuc2003} listed a velocity of $27.9$\kms\ for 
NLTT~3080 close to our own measurement of $28.6$\kms. The standard deviation 
of the six measurements ($5.1$\kms), is larger than individual errors 
($\sim 1$\kms), and is probably typical of the limited stability of the FORS 
velocity scale. \citet{sal2004} measured a radial velocity $v_r =74\pm7$\kms\ 
for the candidate halo star NLTT~9940 in agreement with our velocity of 
$80.5\pm3.5$\kms\ (standard deviation 7.6\kms). \citet{max1999} measured a 
radial velocity $v_r =108.8\pm0.8$\kms\ for NLTT~33503 with a low probability 
of being variable. We measured $111.5\pm3.7$\kms\ (measurement and systematic 
errors added as above) revealing only a small systematic difference. Finally, 
\cite{kaw2012} measured the radial velocity of NLTT~23966 with the X-shooter 
spectrograph and obtained $v_r =66.4\pm5.0$\kms\ in agreement with the present 
FORS measurement of $58.4\pm2.7$\kms\ (standard deviation 7.1\kms). On average, 
our measurements differ by only $\sim 4$\kms\ from published measurements. The differences are comparable to systematic errors of $\sim 5$\kms.

Similarly, \citet{sal2004} measured $v_r =81\pm7$\kms\ for NLTT~5543 while we 
measured $90.3\pm5.3$\kms\ using the H$\beta$ line showing similar accuracy.

\begin{figure}
\includegraphics[width=\columnwidth]{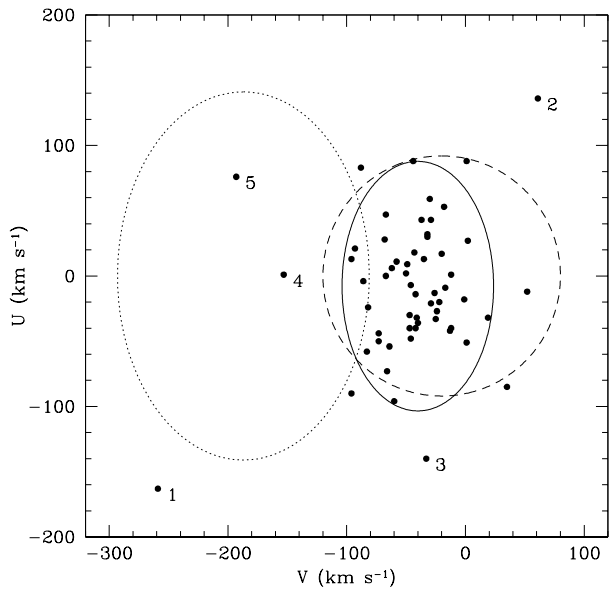}
\caption{The $U$ and $V$ velocities and $2\sigma$ locus (full line) of the NLTT 
white dwarfs compared to the $2\sigma$ thick-disc (dashed line)
and 1$\sigma$ halo ellipses (dotted line) from \citet{chi2000}. Stars of interest
are marked: (1) NLTT~11748, (2) NLTT~13015, (3) NLTT~17486, (4) NLTT~31473, and (5) NLTT~33503.
\label{fig_kinematics}}
\end{figure}

Figure~\ref{fig_kinematics} shows the kinematical properties of our sample in the $U$ versus $V$ Galactic velocity components.
We calculated the velocities following \citet{joh1987}. The effect of the gravitational redshift was removed from the 
radial velocity measurements.
The UVW velocity distributions excluding five peculiar objects (see Fig.~\ref{fig_kinematics}) follow expected averages and standard deviations of an old population: 
$(\bar{U},\bar{V},\bar{W})=(-7.8, -40.1, -5.9)$\kms, 
and $(\sigma_{U},\sigma_{V},\sigma_{W})=(42.8,31.9,27.4)$\kms.  The average $V$ velocity of the present sample is characteristic
of the lag-velocity of thick-disc white dwarfs \citep{pau2006}, but it is larger than
estimated by \citet{chi2000} for thick-disc stars. On the other hand, the velocity dispersions are intermediate to those
of thin- and thick-disc white dwarfs \citep{pau2006}. The present NLTT sample is 
possibly a mix of old thin- and thick-disc populations. 
Three objects clearly belong to the halo: NLTT~11748, NLTT~31473, and NLTT~33503. NLTT~33503 is a newly identified, young halo white dwarf ($\tau_{\rm cool} = 170$ Myr).
We calculated a mass of $0.51\pm0.02\,M_\odot$ and, following the low-metallicity ($Z=0.0004$) evolutionary models of \citet{gir2000}, 
the likely progenitor is a low-mass star ($M\approx 0.85 \,M_\odot$) with a total lifespan of 11~Gyr.
NLTT~31473 is an older halo white dwarf ($\tau_{\rm cool} = 2.9$ Myr) with a more massive progenitor with a lifespan
of 8~Gyr ($M_i\approx 0.94\,M_\odot$).
Two other objects show extreme kinematical properties: the high-field magnetic white dwarf NLTT~13015 and
the visual double-degenerate NLTT~17486. Both objects are beyond the 2$\sigma$ locus of the thick-disc but
they are clearly not associated to the halo.

\subsection{Comments on selected objects}

\subsubsection{Overlap with other catalogues or surveys}

Our selection is based primarily on the catalogue of \citet{sal2003} but some of the
targets may also be found in the Liverpool-Edinburgh high proper-motion survey \citep{pok2003}.
Most objects may be found in 
Luyten's White Dwarf Catalogues \citep{luy1970,luy1977}; these objects were considered probable
white dwarfs based on their reduced proper-motion but, in most cases, they were awaiting spectroscopic
confirmation. Only ten objects were not listed in Luyten's White Dwarf Catalogues.

A few objects from our sample have been studied previously.
\citet{koe2009} obtained VLT/UVES spectra of NLTT 3080 and 
measured (\teff,\,$\log{g}) = (6515\pm7\, {\rm K},\, 7.04\pm0.01)$. 
However, by combining $V=15.26$ from \citet{egg1965a} and the 2MASS $J$, i.e., $V-J=0.75\pm0.04$, we obtain
a temperature of $7000\pm150$ K in agreement with our spectroscopic determinations
(Table~\ref{tbl_atm}).

\citet{lam2000} listed NLTT~5814 as a DA3.5 white dwarf (MCT 0142$-$3026).

\citet{lim2010} measured (\teff,$\, \log{g}) = (10690\pm165\, {\rm K},\, 8.09\pm0.06)$
for NLTT~6559 (KUV01552$-$0703). We obtained a significantly lower surface gravity measurement.

\citet{sub2008} listed NLTT~13532 as a DC white dwarf. However, assuming a hydrogen-rich composition they estimated \teff$\, = 5330\pm146$
based on optical VRI and 2MASS photometric measurements. 
We classified NLTT~13532 as a DA white dwarf based on the detection of H$\alpha$ in FORS spectra.

\citet{rei2005} classified the object NLTT~14558 as a DC white dwarf, although a weak H$\alpha$ line is apparent in their 
spectrum. Based on a photometric temperature estimate, \citet{kaw2004} estimated a relatively high tangential velocity.
We classified NLTT~14558 as a DA white dwarf based on the detection of H$\beta$ in FORS spectra.
The Galactic velocity vectors of NLTT~14558 are typical of the present sample.

\citet{hin1974} classified NLTT~15957 as a DA ``weak'' owing to the weak Balmer lines.

\citet{rei2005} classified NLTT~17486 as a DA7 white dwarf. Based on photometric measurements, 
\citet{sub2008} estimated an
effective temperature of \teff$\, = 9900\pm440$ K. Also, \citet{kaw2004} estimated a high tangential velocity, and,
as noted above (Section 3.4), this object has peculiar kinematics.

NLTT 28493 was classified as a cool DA white dwarf by \citet{kil2006}.
They estimated an effective temperature of \teff$\,= 6516$ K using SDSS
photometry. \citet{car2006} also showed this star to be a cool DA white dwarf.

For NLTT~33503, \citet{koe2009} measured $(T_{\rm eff},\, \log{g}) = (16152\pm25\, {\rm K},\, 7.720\pm0.005)$.

\citet{sub2007} classified NLTT~33669 as a DA white dwarf and estimated $T_{\rm eff} = 6083\pm201$ based on photometry.

\citet{gre1984} classified NLTT~51252 as a DA white dwarf.

NLTT 56805 was classified as a cool DA white dwarf by
\citet{ven2003}. They estimated an effective temperature of $4700\pm300$ K, and
showed that it has a low surface gravity ($\log{g} \approx 7$). Our new
spectrum along with the SDSS $ugriz$ photometry confirm the low temperature. 
The recent trigonometric parallax measurement of $\pi = 44.9\pm2.0$ mas (or $d=22.3\pm1.0$ pc) from
\citet{lep2009} confirms the low surface gravity of the object: assuming
$T_{\rm eff} = 5000$ K and an absolute magnitude of $M_V = 14.3\pm0.1$, we obtain
a low surface gravity of $\log{g} = 7.4\pm0.1$. 

PM~J11480$-$4523 was selected from the sample of high proper-motion
stars of \citet{lep2005}.

\begin{figure*}
\centering
\includegraphics[width=1.0\columnwidth]{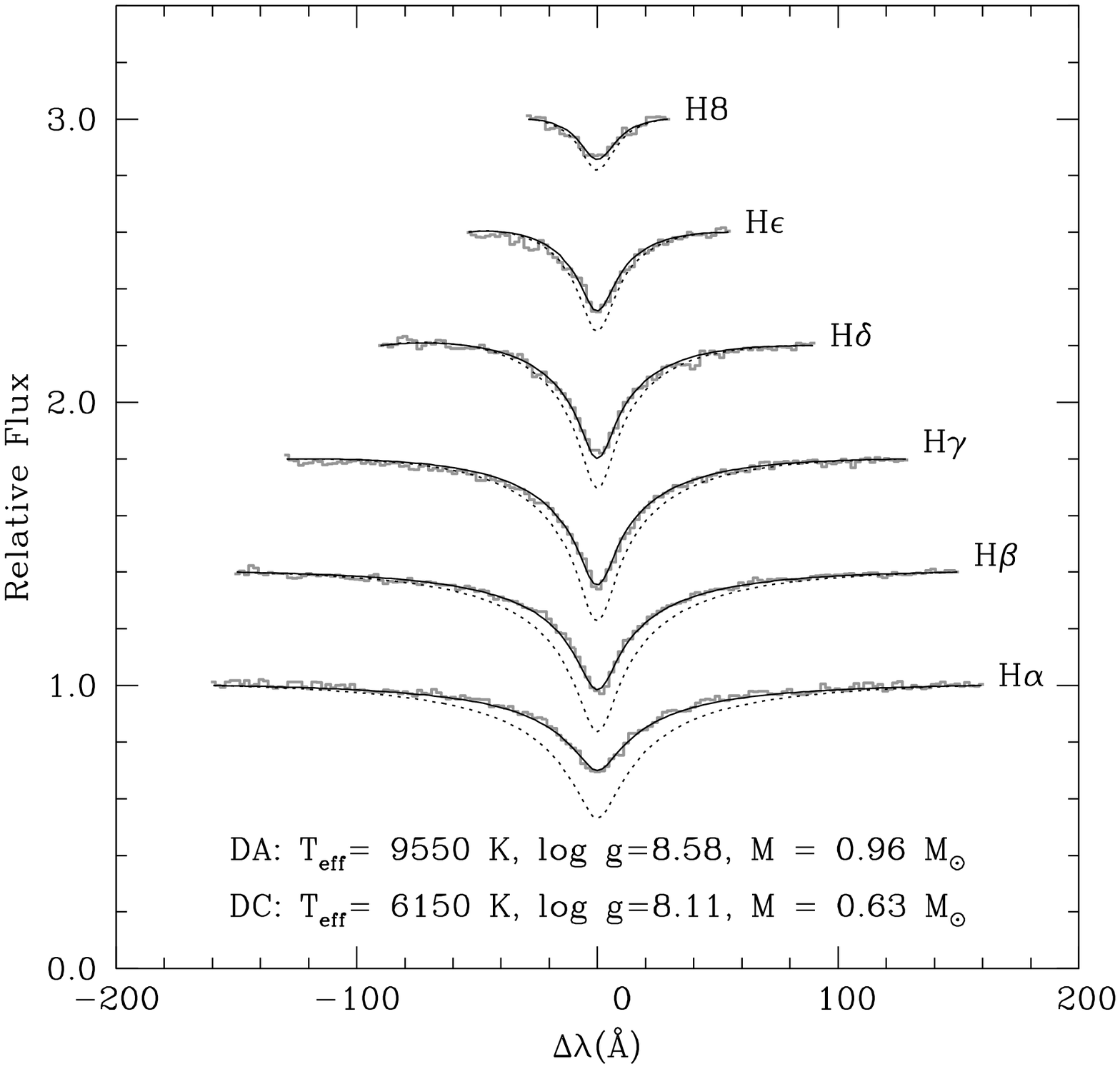}%
\includegraphics[width=1.0\columnwidth]{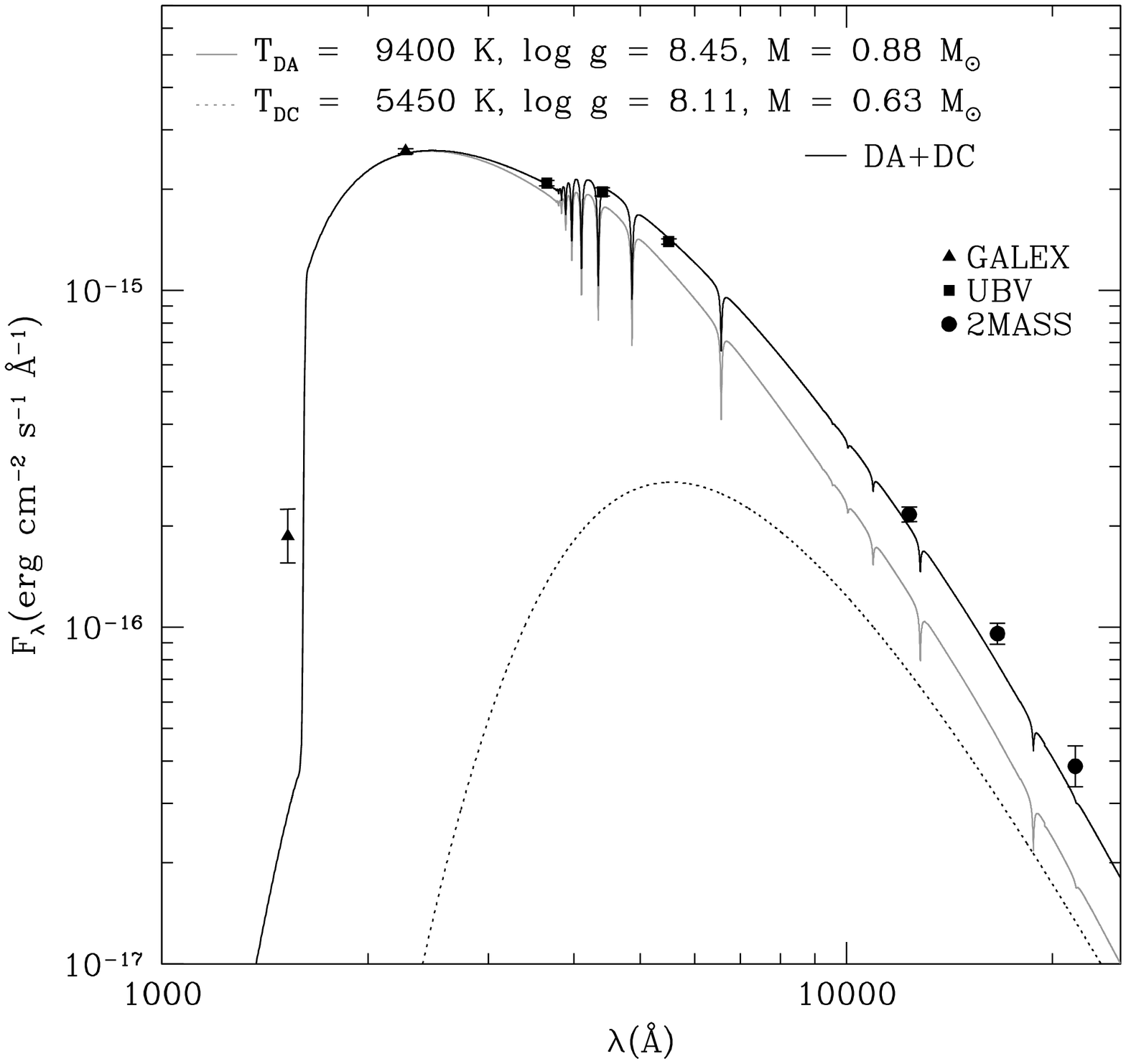}
\caption{{\it Left:} Balmer line profiles of NLTT~21913 obtained at CTIO (grey lines)
compared to the best fitting model spectrum (full lines) composed of a hydrogen-rich 
component at $T_{\rm eff} = 9550$ K, $\log{g} = 8.58$ (dotted lines) that 
is diluted by the helium-rich component at $T_{\rm eff} = 6150$ K, 
$\log{g} = 8.11$.  {\it Right:} Photometry of NLTT~21913 (GALEX 
$FUV,NUV = 20.995\pm0.200,\ 17.278\pm0.018$; 
$U,B,V = 15.78\pm0.02,\ 16.32\pm0.03,\ 16.04\pm0.02$ \citep{kil1997}; 
2MASS $J,H,K = 15.401\pm0.055,\ 15.183\pm0.077,\ 15.111\pm0.152$) compared to 
the best-fitting spectral energy distribution (full line) showing the contributing fluxes 
from the hydrogen- and helium-rich components (grey and dotted lines, respectively).}
\label{fig_21913}
\end{figure*}

\subsubsection{DAZ white dwarfs}

Six objects are classified as DAZ white dwarfs. \citet{kaw2011,kaw2012} already discussed the cases
of NLTT~6390, 10480, and 11393.  Briefly, they measured calcium abundances ranging from
$\log{\rm Ca/H} \approx -10.0$ to $-10.5$, and, as mentioned above, measured an averaged surface
field of $\sim500$ kG in NLTT~10480. The spectrum of NLTT~6390 also showed iron and magnesium lines.
Our survey uncovered two new DAZ white dwarfs, NLTT~888 and 7547, and new spectra of NLTT~25792 helped
reclassify this object as a DAZ white dwarf.
\citet{kil1997} had originally classified
NLTT~25792 (EC~10542$-$2236) as a sdB+, that is a hot sdB star which shows the Ca~K line.
\citet{kaw2011x} and \citet{gia2011} independently identified the star as a cool DAZ white dwarf.
\citet{gia2011} also measured a calcium abundance $\log{\rm Ca/H} =-8.3$ at \teff$\,=7910$ K. Their
abundance is somewhat lower than estimated in the present work, and both measurements are lower
than estimated by \citet{kaw2011x} because of their higher temperature estimate.
Our new measurement of the calcium abundance and effective temperature, $\log{\rm Ca/H} =-7.9\pm0.2$ and \teff$\,=7921$,
as well as the measurements of \citet{gia2011} 
place NLTT~25792 above the observed calcium abundance trend \citep[see][]{kaw2011x}. 

The calcium abundances in the atmospheres of NLTT~888 and  7547 are $\log{\rm Ca/H} = -10.65\pm0.15$ and
$-10.05\pm0.15$, respectively. 

\subsubsection{Double degenerates and common proper-motion binaries}

Several objects are paired with main-sequence or degenerate companions.

The white dwarf NLTT~4615 is in a common 
proper-motion binary with the dM star NLTT~4616 \citep{sil2001}. They 
estimated for the white dwarf \teff$\,=6037$\,K based on the $V-I$ index.
\citet{sch2012} list NLTT~4615 as a ``former'' Hyades DC white dwarf. 
\citet{gar2011} correctly identified it as a DA white dwarf and measured 
\teff$\,=5950$\,K based on $VJHK$ colours, or $\sim$170 K hotter than our 
spectroscopic measurement. 

\citet{sil2001} listed NLTT~7051 as a DA white dwarf and estimated 
\teff$\,=5709$ K. The white dwarf is in a common proper-motion binary with the 
dM3 star NLTT~7055.

\citet{gia2011} independently identified NLTT~21913 as an unresolved DA+DC 
pair. They found that the binary is composed of a massive DA component with 
(\teff,$\, \log{g}) = (10000\, {\rm K},\, 8.91)$ and a cool DC white dwarf with 
(\teff,$\, \log{g}) = (5600\, {\rm K},\, 8.00)$. The average of two radial 
velocity measurements obtained with FORS ($v_r=95.9\pm6.0$\kms) differs 
significantly from the velocity measured by \citet{zuc2003} with the Keck HIRES, $v_r=130.2$\kms, from 
which we conclude that NLTT~21913  is a close DA+DC binary. Additional velocity 
measurements should help constrain the period and DC mass function. We
fitted the observed Balmer lines with two sets of model spectra using $\chi^2$
minimization techniques. In each fitting procedure, we varied \teff\
and $\log{g}$ for the H-rich white dwarf and only the \teff\ for the He-rich
white dwarf at a given mass. We calculated a set of He-rich spectra at 
different masses (0.5, 0.6, 0.7 and 0.8 $M_\odot$) and the fitting procedure
was repeated for each given mass. We took note of the best $\chi^2$ for each
fit and calculated the best mass and temperature for the DC with the 
corresponding \teff\ and $\log{g}$ for the DA white dwarf. 
Figure~\ref{fig_21913} (left panel) shows the best fitting model spectra 
(DA: $T_{\rm eff} = 9550,\ \log{g} = 8.58,\ M=0.96\ M_\odot$, 
DC: $T_{\rm eff} = 6150,\ \log{g} = 8.11,\ M=0.63\ M_\odot$) 
compared to the observed Balmer lines of NLTT~21913. Similarly, we fitted
the spectral energy distribution of NLTT~21913 using the two sets of model
spectra. First, we calculated synthetic magnitudes ($GALEX$ $FUV$, $NUV$,
$UBV$, and 2MASS $JHK$) using our spectra. We then fitted the measured
magnitudes $GALEX$ $NUV$, $UBV$ and 2MASS $J$ magnitudes with the synthetic 
magnitudes assuming a mass of $0.63\ M_\odot$ for the DC white dwarf and
varying both effective temperatures and the surface gravity of the DA white
dwarf. Figure~\ref{fig_21913} (right panel) shows the measured magnitudes
of NLTT~21913 compared to the best fitting combined spectrum and the DA
and DC component spectra. Both methods uncovered a massive DA component.

\citet{egg1965b} classified NLTT~17662 (EGGR~171) as a DA white dwarf. As discussed earlier (Section 2) 
we uncovered in the FORS acquisition image a close companion to this object.
Both stars were acquired during spectroscopic observations with the bright component (A) centred on the
slit and the fainter component (B) off-centre. We measured \teff\ and $\log{g}$ of component A using the EFOSC spectrum obtained as part of
the P82 programme; component B was left out of the slit. Next, we calculated the absolute $R$ magnitude ($M_R$) of component A, and, using
apparent magnitudes (Table~\ref{tbl_mag}) we calculated 
$M_R=14.3\pm0.3$ mag for component B. Finally, we fitted the H$\alpha$ line profile in the \teff\ and $\log{g}$  plane and constrained the range
of solutions using the calculated $M_R$. The resulting parameters are listed in Table~\ref{tbl_atm}.
The surface gravity or mass measurements are strongly correlated because of the requirement of a constant flux ratio.
Component A is hotter and younger than component B. 
Using the evolutionary models of \citet{sch1992} at $Z=0.019$ and keeping the total age within 10 Gyr, 
the cooling age differential ($\Delta \tau_{\rm age}\approx 1.4-2.3$ Gyr) implies an initial
mass range $M_i(A) \approx 1.1-2.0\,M_\odot$ for component A while the cooling age of component B limits the initial mass to $M_i(B) \ga 1.6\,M_\odot$. 
Incidentally, the surface 
gravity measurements suggest that the white dwarf NLTT~17662B is only $0.05-0.14\,M_\odot$ more massive than its companion NLTT~17662A,
in agreement with the small gravitational redshift offset between the two stars ($\gamma_B-\gamma_A\approx 6$\kms). 
Applying additional constraints stemming from the initial-to-final mass relations of \citet{kal2008}, the final mass differential 
implies an initial mass differential of $\approx0.4-1.3\,M_\odot$. Hence, the initial, correlated mass ranges are $M_i(A) \approx 1.2-1.8\,M_\odot$
and $M_i(B) \approx 1.6-3.1\,M_\odot$.
Further spectroscopic observations, such as a complete Balmer line spectrum of component B, should help refine these estimates. 

The white dwarf NLTT 29967 is
in a common proper-motion binary with the K-type dwarf HIP 59519 \citep{gou2004}. The Hipparcos parallax,
$\pi = 22.18\pm1.49$ mas, corresponds to a distance of
$d = 45.1^{+3.2}_{-2.8}$ pc \citep{van2007} in agreement with our distance estimate
of $d=41\pm8$ pc that was obtained using the calculated distance modulus $m-M$.
Conversely, by constraining the distance to $45.1^{+3.2}_{-2.8}$ pc and applying the mass-radius relation 
we constrain the surface gravity to $\log{g} = 7.92^{+0.14}_{-0.17}$ 
and the mass to $M = 0.54\pm0.09\ M_\odot$.
\citet{gar2011} measured $(T_{\rm eff},\, \log{g}) = (6180\pm220\, {\rm K},\, 7.26\pm0.45)$
using Balmer line profiles,
but $T_{\rm eff}=5620\pm160$ using photometric measurements alone.
Our spectroscopic results confirm the higher temperature estimate.

\subsubsection{Halo white dwarfs}

Our kinematical study offered clues to the origin (total age) of some of the objects.

\citet{opp2001} listed NLTT~5543 as a candidate halo white dwarf and
their spectroscopy revealed the presence of the H$\alpha$ line (DA class). \citet{ber2005} estimated
an effective temperature of \teff$\, = 7000\pm180$ K using optical and infrared
photometry. We measured \teff$\, = 6620\pm160$ by fitting the Balmer lines with the surface
gravity set to $8.00\pm0.24$. The effective temperature corresponding to the colour index 
$V-J=0.85\pm0.11$ is \teff$\, = 6640^{+390}_{-340}$~K.
The DA NLTT~9940 is the second halo white dwarf candidate proposed by \citet{opp2001} that is covered in our sample. 
\citet{ber2005} estimated \teff$\,= 5750\pm190$ K based on photometry alone.
We measured \teff$\, = 5660\pm270$ by fitting the Balmer lines, and again setting the surface
gravity at $8.00\pm0.24$. The temperature obtained by fitting the SDSS $ugriz$ colours is $5510\pm30$~K
in agreement with the spectroscopic temperature. Both optical-infrared indices ($V-J$ and $R-J$) indicate a lower temperature of
5000~K possibly revealing an error in
the 2MASS $J$ band measurement: \citet{kil2010} measured $J=16.20$ and the optical-infrared
index $V-J=1.13$ corresponds to a temperature of $\approx$5800~K in agreement with optical temperatures.
We confirm that 
neither object, NLTT~5543 or NLTT~9940, is a genuine halo white dwarf but, instead, they are likely members of
the thick-disc \citep[see also][]{sal2004}.

We originally presented evidence of the peculiar Galactic velocity components of NLTT~11748 \citep{kaw2009,kaw2010}. 
Its halo membership also allowed \citet{kaw2009} to constrain the mass of the progenitor ($M_i\approx 0.9\,M_\odot$).
The DA white dwarfs NLTT~31473 and NLTT~33503 are two new candidates for the halo membership based on their Galactic velocity
components (Fig.~\ref{fig_kinematics}).

\section{Summary and Discussion}

We have conducted a spectropolarimetric survey of 58 DA white dwarfs. 
We report the discovery of the low-field magnetic white dwarf NLTT~2219 
($B_l=-97$ kG) and of the likely extremely low-field white dwarf NLTT~347.
Our survey also led to the identification of the magnetic DAZ white dwarf
NLTT~10480 \citep{kaw2011}. Spectropolarimetric series of the high-field
white dwarf NLTT~12758 revealed short-period ($\la 1$~hr) polarity variations characteristic
of a rotating dipole. We also noted radial velocity variations and a strong contrast
in the strengths of the $\pi$ and $\sigma$ components suggesting that NLTT~12758 is
a spectroscopic binary. Further work on this object is awaiting detailed Zeeman
modelling and additional spectroscopic data to establish the periodicity of the
radial velocity and field variations.
The time series  of NLTT~13015 also showed possible variations 
and the surface-averaged field spread from 6 to 7.5 MG.
The precision of H$\alpha$ velocity measurements also allowed us to identify potential
close double degenerate stars such as the DA+DC NLTT~21913 and the single-lined DA white dwarfs
NLTT~6559 and NLTT~31473. Incidentally, NLTT~31473 is also a member of the Galactic halo. 
\citet{max1999} used their radial velocity survey to determine the fraction of 
double degenerate systems among white dwarfs to be between $\sim 2\%$ and 19\%.
Including the likely double-degenerate NLTT~12758, we estimate the fraction of close double degenerates at $7\pm3$\%.

We exploited the present data to infer stellar (mass, age) and kinematical properties of this
sample of relatively old high-proper motion white dwarfs. For instance, we identified two new halo
white dwarf candidates (NLTT~31473 and NLTT~33503) among a sample of thin- or thick-disc
white dwarfs. 

The atmospheric parameters (\teff\ and $\log{g}$) obtained using spectroscopic
and photometric data are in good agreement. The SDSS $ugriz$ photometry allowed us for the first time to assess
the effect of Ly$\alpha$ extended-wing opacities \citep{kow2006} on the SDSS $u-g$ colour index of cool
DA white dwarfs.

\begin{figure}
\centering
\includegraphics[width=1.0\columnwidth]{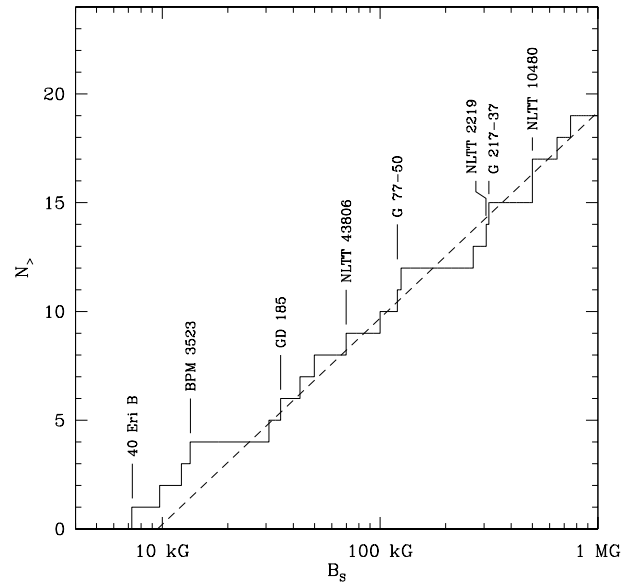}
\caption{Cumulative distribution of low-field measurements ($B_{\rm S} \la 1$MG)
in white dwarfs, where $N_>$ is the number objects with a field larger than a 
surface average field $B_{\rm S}$. Notable objects from Table~\ref{tbl_low} 
are marked. With the exception of a slight enhancement near 10 kG, the observed 
distribution is nearly linear implying a constant incidence of magnetic field 
per decade.}
\label{fig_cum}
\end{figure}

\begin{figure}
\centering
\includegraphics[width=1.0\columnwidth]{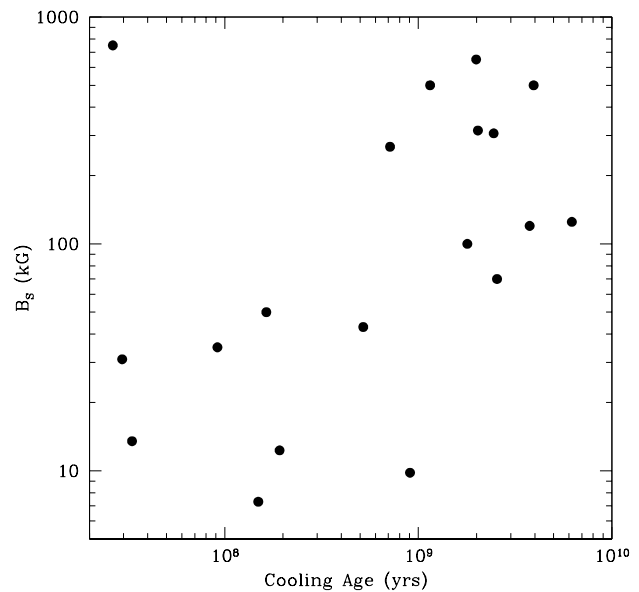}
\caption{Magnetic field strength versus cooling age for low-field white dwarfs
(Table~\ref{tbl_low}).}
\label{fig_mag_age}
\end{figure}

The incidence of magnetism in white dwarfs has been reported to be as low
as 4\% \citep{sch1995} when taking into account all magnetic field strengths,
or $\sim 1$\% per decade interval.
However, it was reported to be 
as high as 25\% when considering only the very low-field white dwarfs
\citep{azn2004}. \citet{azn2004} detected kG fields in three white dwarfs
out of total 12, and when they extended their survey to an additional
10 objects they only reported one possible candidate \citep{jor2007}.
The combined survey hence reduced the incidence of kG fields in white dwarfs
to 11 - 15\%. \citet{kaw2007} studied the incidence of magnetism in the Solar
neighbourhood and found that $21\%\pm8\%$ of white dwarfs are
magnetic. In the present survey of nearly 60 objects we uncovered three new 
low-field white dwarfs
with strengths of the order of $\sim10$ to $\sim10^2$ kG implying a field incidence 
of $5\pm2$\%, or $\sim 1-2$\% per decade interval. The incidence observed
in the present sample is similar to that observed by \citet{sch1995}.

The exact fraction remains uncertain because of inhomogeneous sampling and
methodology.
Figure~\ref{fig_cum} shows the cumulative distribution of low-field white dwarfs as 
a function of the field logarithm. The linear relation corresponds to a nearly 
flat distribution and is evidence that fields are distributed randomly rather 
than following a distribution pattern as found in Ap stars \citep{aur2007}. 
Fields generated by a dynamo involving a common-envelope phase 
\citep[see][]{tou2008,pot2010,nor2011} with a body varying in size, hence disposable 
energy, such as an asteroid, a planet, a brown dwarf, or a low-mass star are likely
to vary in intensity as well. The random nature of the size of the body 
involved in the common-envelope phase should also result in random field 
intensity. Once the field is frozen into the stellar body, the decay time-scale
is of the order of $10^9 - 10^{10}$ years \citep{mus1995}, i.e., comparable
to the white dwarf cooling age. 
On the other hand, the Ap stars are still 
likely progenitors of high-field ($\ga 10^7$ G) white dwarfs.

Are two independent formation channels possible? \citet{tou2008} 
argues that the absence of magnetic white dwarfs in non-interacting
pairs implies that all magnetic white dwarfs are the products of
interaction or merger.  This syllogism does not exclude other possible channels
if it can be shown that the failure to identify magnetic white dwarfs in
non-interacting binaries is the result of a selection effect for this particular channel. For example, 
\citet{car2002} found that $\sim 40$\% of Ap
stars are in binaries. The inferred distribution of mass ratios implies
that close to half of the progeny of Ap stars, i.e., the high-field magnetic white dwarfs, would be
paired with luminous companions (G- to A-type) hence escaping detection, although the remainder would be paired with M- to K-type stars and should be detectable, at least during
the early cooling stages. As noted by \citet{tou2008}, late-type companions remain elusive,
but it remains to be shown that early-type companions are missing as well.

The stellar ages in this sample vary from 20 Myr to 5 Gyr. The distribution of 
field strengths as a function of temperatures possibly reveals strong
selections effects in the sample. Figure~\ref{fig_mag_age} shows that
all stars from Table~\ref{tbl_low} and with a field 
$\le 50$ kG are relatively young stars with cooling ages $\la 10^9$ 
years. Conversely, all stars save two with a field stronger than 50 kG are
older stars with cooling ages in excess of $\sim 10^9$ years. Deep, narrow 
spectral lines are reliable field tracers but are also lacking in cool white 
dwarfs unless heavy element lines are present such as in the magnetic DAZ 
NLTT~43806 \citep{zuc2011} or NLTT~10480 \citep{kaw2011}. On the other hand, an explanation for the paucity 
of intermediate fields (100 kG$-$1 MG) in younger white dwarfs is not readily 
available. Although these are the field and temperature ranges targeted in 
most spectropolarimetric surveys \citep[see, e.g.,][]{azn2004}, a larger 
survey may yet uncover the missing objects.

\begin{table}
\centering
\begin{minipage}{\columnwidth}
\caption{Low-field ($B_{\rm S} \la 1$ MG) white dwarfs. \label{tbl_low}}
\renewcommand{\footnoterule}{\vspace*{-15pt}}
\centering
\begin{tabular}{lrcccc}
\hline\hline
Name & \teff\  & Age \footnote{Calculated assuming $\log{g}=8$.} & $B_{\rm S}$  & Method \footnote{Original measurement from spectropolarimetry ($B_{\rm l}$) or spectroscopy ($B_{\rm S}$). The equivalent
$B_{\rm S}$ is estimated at $i=57^\circ$, i.e., $B_{\rm S}\approx 3.16\times B_{\rm l}$.} & 
 Ref. \footnote{
 References: 1 - \citet{fab2000}; 2 - \citet{azn2004}; 3 - \citet{koe1998}; 4 - \citet{zuc2011};
 5 - \citet{ber1997}; 6 - \citet{far2011}; 7 - \citet{put1997}; 8 - \citet{sch1994}; 9 - this work;
 10 - \citet{max2000}; 11 - \citet{kaw2011}; 12 - \citet{duf2006}; 13 - \citet{wes2001}.} \\
              &  (K)       & (Gyr)& (kG) &             &     \\
\hline
40~Eri~B      & 16500      & 0.15 & 7.3  & $B_{\rm l}$ &  1  \\      
LTT~9857      &  8660      & 0.91 & 9.8  & $B_{\rm l}$ &  2  \\      
LTT~4099      & 15280      & 0.19 & 12.3 & $B_{\rm l}$ &  2  \\      
BPM~3523      & 23450      & 0.03 & 13.5 & $B_{\rm l}$ &  2  \\      
2329$-$291    & 24000      & 0.03 & 31.  & $B_{\rm S}$ &  3  \\      
1531$-$022    & 18850      & 0.09 & 35.  & $B_{\rm S}$ &  3  \\      
2105$-$820    & 10760      & 0.52 & 43.  & $B_{\rm S}$ &  3  \\      
2039$-$682    & 16050      & 0.16 & 50.  & $B_{\rm S}$ &  3  \\      
NLTT~43806    &  5900      & 2.55 & 70.  & $B_{\rm S}$ &  4  \\      
LHS~5064      &  6680      & 1.79 & 100. & $B_{\rm S}$ &  5  \\      
G~77-50       &  5310      & 3.76 & 120. & $B_{\rm S}$ &  6  \\      
G~234-4       & $\sim$4500 & 6.21 & 125. & $B_{\rm l}$ &  7  \\      
LP~907-37     & $\sim$9500 & 0.71 & 268. & $B_{\rm l}$ &  8  \\      
NLTT~2219     & 5980       & 2.45 & 307. & $B_{\rm l}$ &  9  \\      
G~217-37      & $\sim$6400 & 2.03 & 316. & $B_{\rm l}$ &  8  \\      
1953$-$011    & $\sim$7900 & 1.15 & 500.\footnote{Highest of the two components.} & $B_{\rm S}$ &  10  \\   
NLTT~10480    & $\sim$5250 & 3.94 & 500. & $B_{\rm S}$ &  11  \\      
G~165-7       & 6440       & 1.99 & 650. & $B_{\rm S}$ &  12  \\      
LB~8915      & $\sim$24500 & 0.03 & 750. & $B_{\rm S}$ &  13  \\      
\hline
\end{tabular}
\end{minipage}
\end{table}

\section*{Acknowledgments}

S.V. and A.K. acknowledge support from the Grant Agency of the Czech
Republic (GA \v{C}R P209/10/0967 and GA \v{C}R P209/12/0217). A.K.
also acknowledges support from the Centre for Theoretical Astrophysics
(LC06014). This work was also supported by the project RVO:67985815.

This publication makes use of data products from the Two Micron All Sky Survey, 
which is a joint project of the University of Massachusetts and the Infrared Processing 
and Analysis Center/California Institute of Technology, funded by the National Aeronautics 
and Space Administration and the National Science Foundation.
This publication also makes use of SDSS spectroscopic and photometric data. 
Funding for SDSS-III has been provided by the Alfred P. Sloan Foundation, the Participating 
Institutions, the National Science Foundation, and the U.S. Department of Energy Office of Science. 
SDSS-III is managed by the Astrophysical Research Consortium for the Participating Institutions 
of the SDSS-III Collaboration.

\appendix
\section{Photometric data}

Images obtained during our P82 and P83 EFOSC programmes were obtained with either the $V$ filter or the $R$ 
filter. We employed a subset of stars also listed in the SDSS photometric survey to calibrate the EFOSC images.
First, we computed synthetic relations between the SDSS $g$ and $r$ and the Johnson $V$ and $R$ magnitudes. Using a sequence
of models at 0.6\,$M_\odot$ and \teff$=5$ to $20\times10^3$ K we obtained
\begin{displaymath}
V = r +0.46\times(g-r)-0.069,
\end{displaymath}
\begin{displaymath}
R = r -0.13\times(g-r)-0.191,
\end{displaymath}
and converted the set of SDSS $gr$ magnitudes from Table~\ref{tbl_sdss} into Johnson $V$ and $R$ magnitudes.
Next, we measured the count rates $c$ from the acquisition images and calculated the instrumental magnitudes:
\begin{displaymath}
V\ (R)_{\rm acq.} = -2.5\log{c}+25.
\end{displaymath}
Finally, simple linear relations were found between $V\ (R)$ and $V\ (R)_{\rm acq.}$ using the SDSS sample:
\begin{displaymath}
V\ (R)_{\rm EFOSC} \equiv V\ (R) = V\ (R)_{\rm acq.} + C_{V(R)} ,
\end{displaymath}
with standard deviations $\approx 0.01$ mag for $V$ and 0.03 mag for $R$. Having estimated the constants $C_{V}$ and $C_{R}$, the calibration was 
applied to the whole EFOSC sample to measure $V$ and $R$ magnitudes.
The uncertainty on the $R$ magnitude of the faint companion to NLTT~17662 is larger at 0.05 mag.
We employed airmass coefficients averaged over the $V$ and $R$ bands, $k_V\approx 0.13$ and $k_R\approx 0.10$  \citep{pat2011}.
The new magnitudes are listed in Table~\ref{tbl_obs} along with published $V$ magnitudes and 2MASS $J$ magnitudes \citep{skr2006}.

\begin{figure}
\centering
\includegraphics[width=1.0\columnwidth]{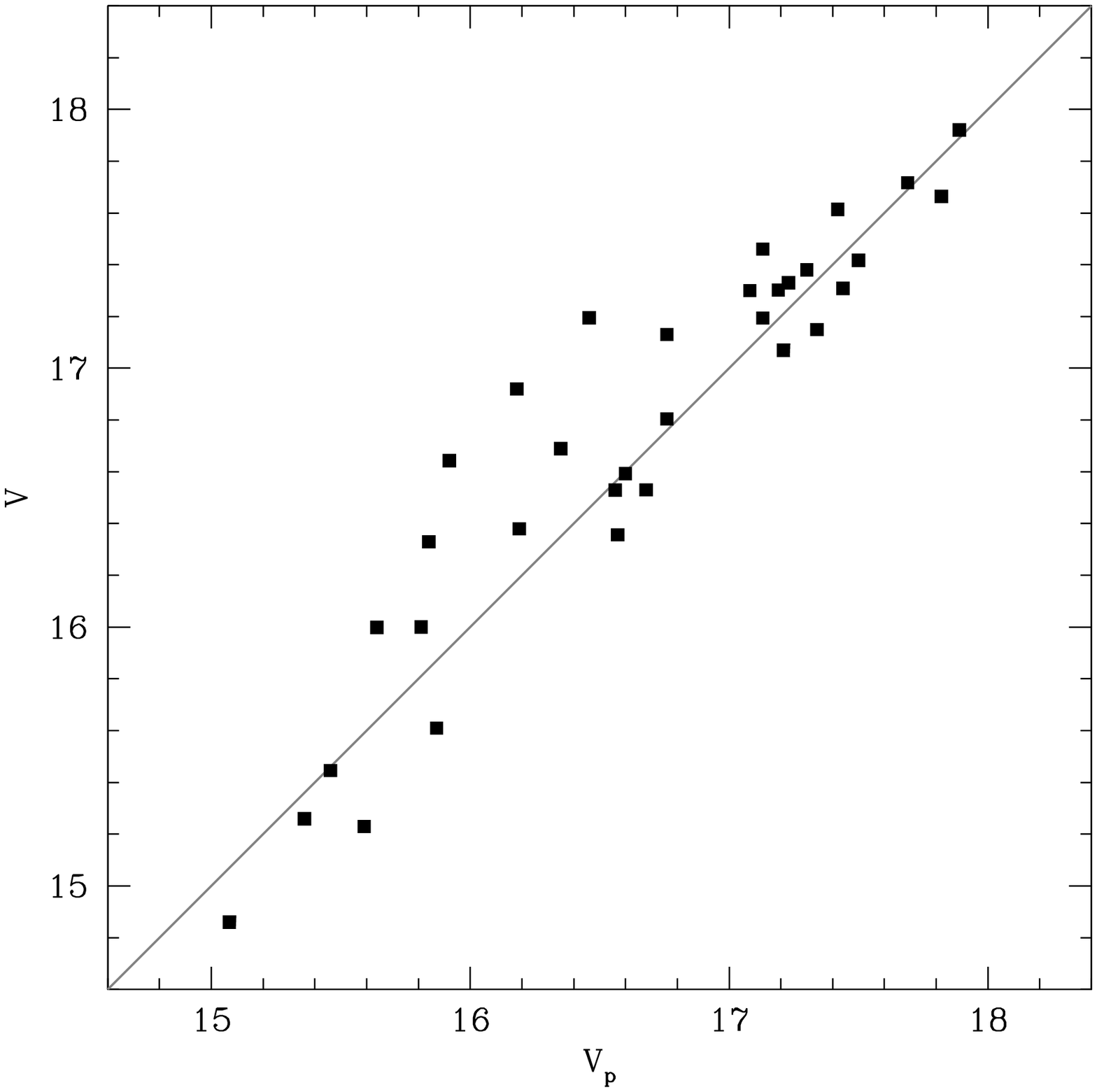}%
\caption{Photographic visual magnitudes ($V_p$) from the rNLTT catalogue of \citet{sal2003} versus calibrated CCD photometry ($V$) from
Table~\ref{tbl_mag}.
}
\label{figvpv}
\end{figure}

The images obtained during our P80 and P84 FORS programmes were obtained in white light. The CCD employed for the P80 observations
was more sensitive in the blue and appropriate to estimate $V$ magnitudes, while the CCD employed for the P84 observations
was appropriate to estimate $R$ magnitudes. We followed the procedure outlined above to calibrate the FORS images. The standard
deviations were 0.04 and 0.03 mag for the $V$ and $R$ calibration stars, respectively. 

\begin{table*}
\begin{minipage}{\textwidth}
\caption{Optical and infrared magnitudes.\label{tbl_mag}}
\centering
\renewcommand{\footnoterule}{\vspace*{-15pt}}
\begin{tabular}{lccccccccc}
\hline\hline
Name & \multicolumn{2}{c}{EFOSC} & & \multicolumn{2}{c}{FORS} & & \multicolumn{2}{c}{Published} & 2MASS  \\
\cline{2-3} \cline{5-6} \cline{8-9}\\
             & $V$       & $R$       & & $V$       & $R$       & & $V\ (\sigma_V)$       & Ref. \footnote{References: 1 - \citet{egg1965a}; 
2 - \citet{smi1997}; 3 - \citet{ber2005}; 4 - \citet{kaw2012}; 5 - \citet{kaw2011}; 6 - \citet{egg1968}; 7 - \citet{sal2003}; 8 - \citet{sub2008}; 
9 - \citet{kil1997}; 10 - \citet{sub2007}; 11 - \citet{cos2006}; 12 - \citet{gre1984}; 13 - \citet{lep2009}; 14 - \citet{lep2005}.} & $J\ (\sigma_J)$  \\
\hline
NLTT~82      &    ... &  ... & &  ... &16.53 & &  ...        &   & 16.120(0.093)\\
NLTT~347     &    ... &  ... & &  ... &16.77 & &  ...        &   & 16.163(0.094)\\
NLTT~888     &  17.669& ...  & &17.659& ...  & &  ...        &   & 16.478(0.104)\\
NLTT~2219    &  17.297& ...  & &17.307& ...  & &  ...        &   & 16.392(0.096)\\
NLTT~2886    &    ... & ...  & &  ... &16.57 & &  ...        &   & 15.901(0.091)\\
NLTT~3080    &    ... & ...  & &  ... & ...  & &  15.26(0.05)& 1 & 14.506(0.033)\\
NLTT~3471    &  16.593& ...  & &  ... &16.43 & &  ...        &   & 16.445(0.123)\\
NLTT~4615    &  17.440& ...  & &  ... &17.09 & &  17.48(0.02)& 2 & 16.461(0.092)\\
NLTT~5503    &   ...  & ...  & &  ... &16.26 & &  ...        &   & 16.276(0.087)\\
NLTT~5543    &  17.137& ...  & &17.127& ...  & &  17.18(0.03)& 3 & 16.296(0.091)\\
NLTT~5814    &   ...  & ...  & &  ... &16.43 & &  ...        &   & 16.531(0.101)\\
NLTT~6004    &   ...  & ...  & &16.329& ...  & &  ...        &   & 15.597(0.060)\\
NLTT~6390    &   ...  & ...  & &17.265& ...  & &  17.33(0.04)& 4 & 16.355(0.128)\\
NLTT~6559    &  16.357& ...  & &  ... &16.24 & &  ...        &   & 16.272(0.106)\\
NLTT~6794    &   ...  & ...  & &  ... &16.99 & &  ...        &   & 16.528(0.113)\\
NLTT~7051    &   ...  &  ... & &  ... &16.47 & &  16.92(0.02)& 2 & 15.536(0.055)\\
NLTT~7547    &  17.946& ...  & &17.910& ...  & &  ...        &   & 16.636(0.133)\\
NLTT~8432    &   ...  & ...  & &  ... &16.91 & &  ...        &   & 16.190(0.091)\\
NLTT~9940    &  17.309& ...  & &  ... &16.97 & &  17.35(0.03)& 3 & 15.819(0.081)\footnote{\citet{kil2010} measured $J=16.20\pm0.04$.} \\  
NLTT~10480   &  17.380& ...  & &  ... & ...  & &  17.49(0.05)& 5 & 16.003(0.078)\\
NLTT~10884   &  17.195& ...  & &  ... & ...  & &  ...        &   & 16.666(0.138)\\
NLTT~11393   &  17.070& ...  & &  ... & ...  & &  17.16(0.04)& 4 & 16.036(0.084)\\
NLTT~11748   &  16.656& ...  & &  ... &16.44 & &  ...        &   & 15.837(0.077)\\
NLTT~12758   &  15.446& ...  & &  ... & ...  & &  15.46(0.05)& 6 & 14.809(0.032)\\
NLTT~12796   &  ...   & ...  & &  ... &16.14 & &  ...        &   & 16.039(0.078)\\
NLTT~13015   &  17.417& ...  & &  ... &17.10 & &  ...        &   & 16.297(0.111)\\
NLTT~13471   &  ...   & ...  & &  ... &16.72 & &  ...        &   & 15.998(0.078)\\
NLTT~13532   &  ...   & ...  & &  ... &16.33 & &  ...        &   & 15.369(0.050)\\
NLTT~13755   &  ...   & ...  & &  ... & ...  & &  16.83(0.25)& 7 & 15.983(0.063)\\
NLTT~14491   &  17.194& ...  & &  ... & ...  & &  ...        &   & 16.315(0.100)\\
NLTT~14558   &  17.717& ...  & &  ... & ...  & &  ...        &   & 16.466(0.120)\\
NLTT~15882   &  ...   & ...  & &  ... &16.85 & &  ...        &   & 16.107(0.080)\\
NLTT~15957   &  16.804& ...  & &  ... & ...  & &  ...        &   & 16.064(0.076)\\
NLTT~17486   &  ...   &15.503& &  ... &15.44 & &  15.61(0.03)& 8 & 15.485(0.063)\\
NLTT~17662A  &   ...  &15.795& &  ... & ...  & &  ...        &   & 15.435(0.056)\\
NLTT~17662B  &  ...   &16.82 & &  ... & ...  & &  ...        &   &  ...        \\
NLTT~17874   &  ...   &16.433& &  ... &16.48 & &  ...        &   & 15.711(0.073)\\
NLTT~21844   &  ...   &16.349& &16.643& ...  & &  ...        &   & 15.989(0.066)\\
NLTT~21913A$+$B &  ...   & ...  & &  ... & ...  & &  16.04(0.02)& 9 & 15.401(0.055)\\
NLTT~23966   &  ...   & ...  & &  ... &16.89 & &  17.13(0.04)& 4 & 16.544(0.112)\\
NLTT~24406   &  ...   &17.121& &  ... &17.02 & &  ...        &   & 16.421(0.107)\\
NLTT~25792   &  ...   & ...  & &15.998& ...  & &  ...        &   & 15.521(0.050)\\
NLTT~28493   &  ...   & ...  & &16.359& ...  & &  16.39(0.03)& 10& 15.543(0.060)\\
NLTT~28730   &  ...   &16.310& &16.529& ...  & &  ...        &   & 16.207(0.082)\\
NLTT~29967   &  ...   &16.711& &  ... & ...  & &  ...        &   & 16.018(0.087)\\
NLTT~31473   &  ...   & ...  & &  ... &16.15 & &  16.53(0.02)& 11& 15.354(0.047)\\
NLTT~31483   &  ...   &17.148& &17.614& ...  & &  ...        &   & 16.592(0.113)\\
NLTT~32785   &  ...   & ...  & &  ... &16.79 & &  ...        &   & 15.833(0.062)\\
NLTT~33503   &  ...   & ...  & &  ... & ...  & &  14.86(0.02)& 9 & 15.172(0.049)\\
NLTT~33669   &  ...   & ...  & &16.689& ...  & &  ...        &   & 15.563(0.051)\\
NLTT~35570   &  ...   & ...  & &  ... &16.71 & &  ...        &   & 16.139(0.092)\\
NLTT~38356   &  ...   & ...  & &17.309& ...  & &  ...        &   & 16.560(0.130)\\
NLTT~50161   &  ...   & ...  & &  ... &16.99 & &  ...        &   & 16.789(0.118)\\
NLTT~51252   &  15.230& ...  & &  ... & ...  & &  15.23(0.05)& 12& 15.606(0.060)\\
NLTT~56045A  &  ...   & ...  & &  ... &16.55 & &  ...        &   & 16.390(0.113)\\
NLTT~56045B  &  ...   & ...  & &  ... &17.64 & &  ...        &   &   ...        \\
\hline
\end{tabular}
\end{minipage}
\end{table*}

\begin{table*}
\begin{minipage}{\textwidth}
\contcaption{}
\centering
\renewcommand{\footnoterule}{\vspace*{-15pt}}
\begin{tabular}{lccccccccc}
\hline\hline
Name & \multicolumn{2}{c}{EFOSC} & & \multicolumn{2}{c}{FORS} & & \multicolumn{2}{c}{Published} & 2MASS  \\
\cline{2-3} \cline{5-6} \cline{8-9}\\
             & $V$       & $R$       & & $V$       & $R$       & & $V\ (\sigma_V)$       & Ref. \footnote{References: 1 - \citet{egg1965a}; 
2 - \citet{smi1997}; 3 - \citet{ber2005}; 4 - \citet{kaw2012}; 5 - \citet{kaw2011}; 6 - \citet{egg1968}; 7 - \citet{sal2003}; 8 - \citet{sub2008}; 
9 - \citet{kil1997}; 10 - \citet{sub2007}; 11 - \citet{cos2006}; 12 - \citet{gre1984}; 13 - \citet{lep2009}; 14 - \citet{lep2005}.} & $J\ (\sigma_J)$  \\
\hline
NLTT~56257   &  ...   & ...  & &  ... &15.46 & &  ...        &   & 15.217(0.035)\\
NLTT~56493   &  ...   & ...  & &  ... &17.05 & &  ...        &   & 15.772(0.063)\\
NLTT~56805   &  15.976& ...  & &15.977& ...  & &  16.05(0.05)& 13& 14.512(0.034)\\
PMJ11480$-$4523 & ... & ...  & &  ... & ...  & &  15.66(0.25)& 14& 14.888(0.039)\\
\hline
\end{tabular}
\end{minipage}
\end{table*}

Figure~\ref{figvpv} shows photographic \citep[$V_p$][]{sal2003} and CCD photometric measurements ($V$) from our sample of stars. The photographic
measurements are on average 0.1 mag brighter than the CCD measurements with a standard deviation of 0.29 mag.

\begin{figure*}
\centering
\includegraphics[width=1.0\columnwidth]{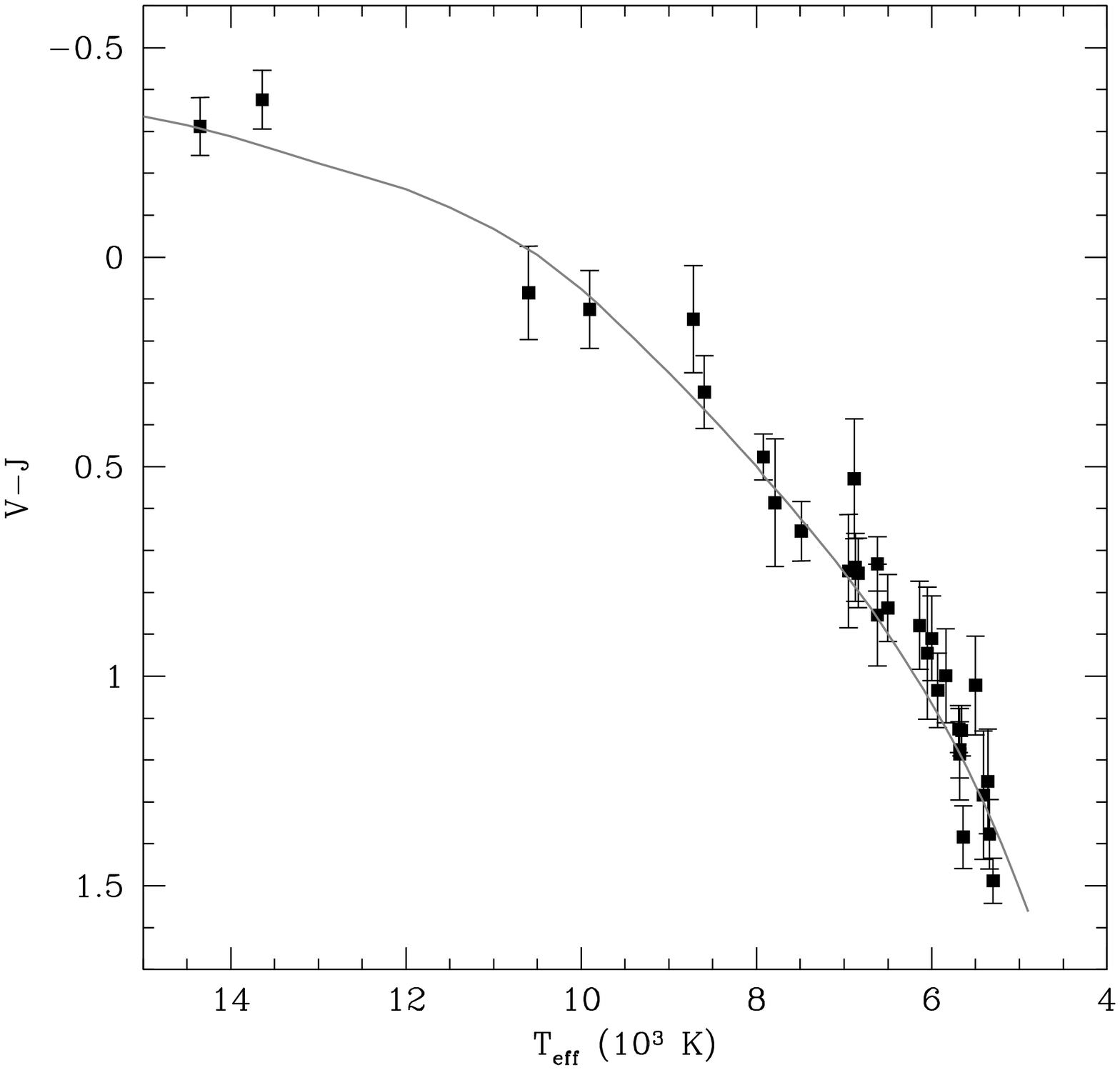}%
\includegraphics[width=1.0\columnwidth]{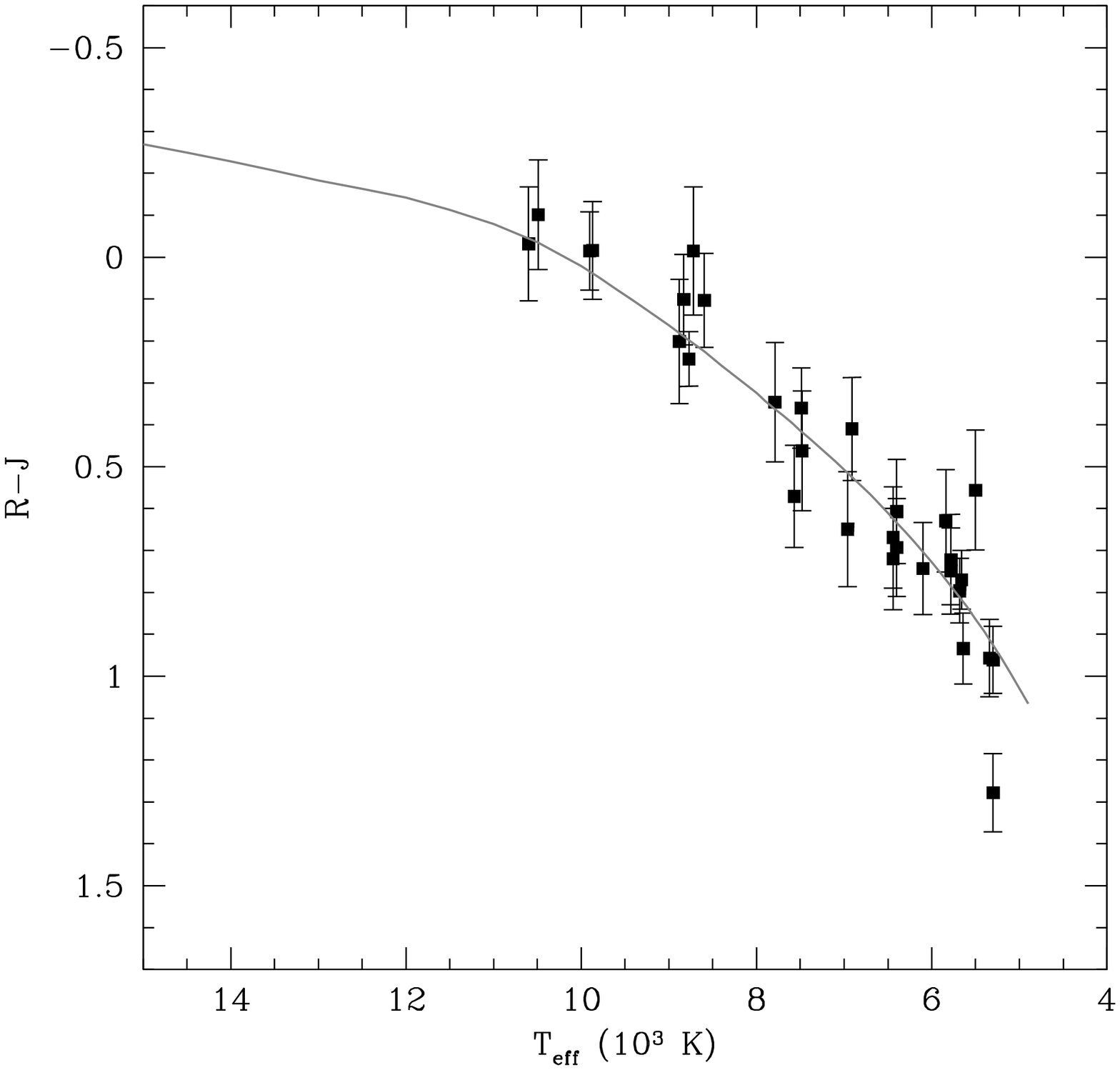}%
\caption{Photometric $V-J$ (left panel) and $R-J$ (right panel) indices as a function of 
effective temperatures measured using Balmer line profiles. The synthetic relations (grey lines)
may be employed to obtain photometric temperature estimates.
}
\label{fig_VRJ}
\end{figure*}

Figure~\ref{fig_VRJ} shows the measured and synthetic $V-J$ and $R-J$ indices 
as a function of temperature. The blended 2MASS $J$ measurements of the resolved binaries NLTT~17662 and NLTT~56045, and
the composite colours of the close binary NLTT~21913 were excluded from the diagrams.
The adopted effective temperature measurements are from Table~\ref{tbl_atm}. Excluding NLTT~11748 \footnote{
The spectral energy distribution of the extremely low-mass white dwarf NLTT~11748 shows the effect of reddening \citep[$E_{B-V}=0.1$,][]{kaw2009}},
the average deviation between predicted and
measured $V-J$ is $+0.046$ mag with a standard deviation of $0.096$ mag, while the
average deviation for the $R-J$ index is $+0.007$ with a standard deviation of $0.119$ mag.
Applying the synthetic relations to the data, again excluding NLTT~11748, we calculated a reduced $\chi^2$ of $\sim 1.2$ in
the $R-J$ diagram and $\sim 1.3$ in the $V-J$ diagram.
The effect of crowded companions or dust is likely to affect the colours of a few of the outliers found in these diagrams.
We noted a discrepancy of $\sim0.4$~mag between the 2MASS $J$ magnitude of NLTT~9940 and the $J=16.20\pm0.04$ measurement of \citet{kil2010}.
Opting for the latter, the revised $V-J$ and $R-J$ indices are in excellent agreement with the optical temperatures (see Section 3.5.4).
We conclude that errors in optical and infrared magnitudes account for the dispersion. However, the $R-J$ index of NLTT~56493 reveals an object
markedly redder than expected suggesting a possible contamination of the 2MASS~$J$ measurement.

Table~\ref{tbl_sdss} lists SDSS photometric data depicted in Figure~\ref{fig_umg_gmr}.

\begin{table*}
\centering
\begin{minipage}{\textwidth}
\caption{SDSS photometry. \label{tbl_sdss}}
\renewcommand{\footnoterule}{\vspace*{-15pt}}
\centering
\begin{tabular}{llccccc}
\hline\hline
NLTT & SDSS & $u$ & $g$ & $r$ & $i$ & $z$ \\
\hline
   82 & J000410.41-034008.5 & $17.449\pm0.010$ & $16.917\pm0.005$ & $16.742\pm0.005$ & $16.708\pm0.006$ & $16.728\pm0.009$  \\
  888 & J001737.86-051649.6 & $19.011\pm0.028$ & $17.983\pm0.006$ & $17.536\pm0.006$ & $17.352\pm0.006$ & $17.344\pm0.017$  \\
 2219 & J004056.24-080912.4 & $18.442\pm0.020$ & $17.575\pm0.005$ & $17.208\pm0.005$ & $17.106\pm0.006$ & $17.069\pm0.012$  \\
 3471 & J010319.70-032501.0 & $17.036\pm0.008$ & $16.646\pm0.004$ & $16.659\pm0.005$ & $16.705\pm0.005$ & $16.812\pm0.010$  \\
 4615 & J012314.84-020926.9 & $18.690\pm0.022$ & $17.729\pm0.005$ & $17.306\pm0.005$ & $17.158\pm0.006$ & $17.148\pm0.015$  \\
 5503 & J013852.65-035650.9 & $16.793\pm0.008$ & $16.351\pm0.004$ & $16.441\pm0.004$ & $16.552\pm0.005$ & $16.730\pm0.009$  \\
 6559 & J015742.08-064847.8 & $16.830\pm0.009$ & $16.363\pm0.004$ & $16.476\pm0.005$ & $16.579\pm0.005$ & $16.839\pm0.011$  \\
 7547 & J021719.98-065628.5 & $19.677\pm0.037$ & $18.323\pm0.007$ & $17.746\pm0.006$ & $17.528\pm0.007$ & $17.440\pm0.014$  \\
 9940 & J030713.90-071506.1 & $18.698\pm0.020$ & $17.635\pm0.005$ & $17.179\pm0.005$ & $16.988\pm0.006$ & $16.937\pm0.014$  \\
17874 & J072704.21+143439.5 & $18.202\pm0.014$ & $17.135\pm0.005$ & $16.692\pm0.005$ & $16.515\pm0.006$ & $16.455\pm0.008$  \\
21844 & J092840.28+184113.6 & $17.095\pm0.008$ & $16.696\pm0.004$ & $16.580\pm0.005$ & $16.574\pm0.005$ & $16.676\pm0.009$  \\
24406 & J102747.63+192824.1 & $17.881\pm0.013$ & $17.387\pm0.005$ & $17.228\pm0.005$ & $17.178\pm0.005$ & $17.254\pm0.012$  \\
25792 & J105638.59-225254.3 & $16.540\pm0.007$ & $16.079\pm0.004$ & $15.974\pm0.004$ & $15.984\pm0.005$ & $16.062\pm0.007$  \\
28493 & J114625.77-013636.9 & $17.119\pm0.008$ & $16.499\pm0.004$ & $16.235\pm0.003$ & $16.152\pm0.004$ & $16.146\pm0.007$  \\
31473 & J124024.15-231743.9 & $17.917\pm0.013$ & $16.814\pm0.004$ & $16.337\pm0.004$ & $16.151\pm0.004$ & $16.084\pm0.007$  \\
31483 & J124030.26+180728.2 & $19.201\pm0.023$ & $17.990\pm0.007$ & $17.416\pm0.006$ & $17.193\pm0.006$ & $17.114\pm0.011$  \\
56493 & J231935.43-022903.3 & $18.980\pm0.024$ & $17.582\pm0.005$ & $16.922\pm0.005$ & $16.653\pm0.006$ & $16.580\pm0.009$  \\
56805 & J232519.88+140339.7 & $18.051\pm0.014$ & $16.461\pm0.004$ & $15.867\pm0.005$ & $15.570\pm0.005$ & $15.405\pm0.006$  \\
\hline
\end{tabular}
\end{minipage}
\end{table*}

\section{Radial velocity measurements}

Table~\ref{tbl_rad_vel} list individual heliocentric velocity measurements using the H$\alpha$ line. Quoted errors are statistical only, i.e.,
follow from the line core fit. We recommend to add 5\kms\ in quadrature to account for accidental shifts in the wavelength scale
from night-to-night. The velocities are measured relative to H$\alpha$ ($\lambda_0=6562.80$\AA) in non-magnetic stars and relative to the
shifted $\pi$ component of H$\alpha$ computed at the appropriate field strength in NLTT~13015 ($\lambda_\pi=6560.43$\AA).
The velocity measurements of NLTT~17662B were corrected by adjusting the telluric lines of the B component with that of the A component because
the spectra were obtained with the A component centred on the slit and the B component significantly off-centre.
We calculated and applied a correction of $+34.4$\kms\ to the spectra of NLTT~17662B.

\begin{table*}
\centering
\begin{minipage}{\textwidth}
\caption{Individual radial velocity measurements (H$\alpha$). \label{tbl_rad_vel}}
\renewcommand{\footnoterule}{\vspace*{-15pt}}
\centering
\begin{tabular}{lcclcclcc}
\hline\hline
 NLTT & BJD          & $v_r$         & NLTT  & BJD          & $v_r$         & NLTT   & BJD          & $v_r$         \\
      & ($2450000+$) & (\kms) &       & ($2450000+$) & (\kms) &        & ($2450000+$) & (\kms) \\
\hline
00082 & 5129.54679 &  $38.9\pm 2.0$ & 12758 & 5127.81251 &  $83.5\pm 2.5$ & 17662A& 5173.80780 &  $43.2\pm 1.0$ \\
00082 & 5129.56148 &  $38.5\pm 4.9$ & 12758 & 5127.82371 &  $86.6\pm 2.5$ & 17662A& 5173.82249 &  $41.2\pm 1.2$ \\
00082 & 5129.58041 &  $40.9\pm 2.9$ & 12758 & 5127.83957 &  $80.8\pm 1.7$ & 17662B& 5173.80780 &  $43.2\pm 3.5$ \\
00082 & 5129.59510 &  $44.8\pm 2.2$ & 12758 & 5127.85088 &  $81.3\pm 3.1$ & 17662B& 5173.82249 &  $53.1\pm 3.3$ \\
00347 & 5129.61463 &  $ 4.4\pm 1.5$ & 12758 & 5159.62116 & $110.2\pm 2.7$ & 17874 & 5177.75553 &  $36.6\pm 3.6$ \\
00347 & 5129.62932 &  $ 5.5\pm 1.8$ & 12758 & 5159.63236 & $122.9\pm10.1$ & 17874 & 5177.77055 &  $35.4\pm 2.0$ \\
00347 & 5129.64781 &  $12.2\pm 1.1$ & 12758 & 5159.65188 & $118.7\pm 5.0$ & 17874 & 5177.79091 &  $37.2\pm 1.3$  \\
00347 & 5129.66248 &  $11.8\pm 1.3$ & 12758 & 5159.66308 & $122.5\pm 4.7$ & 17874 & 5177.80594 &  $37.2\pm 2.8$ \\
02886 & 5119.58441 &  $77.4\pm10.9$ & 12796 & 5168.62568 &  $62.9\pm 2.9$ & 21913A & 5174.82596 &  $98.1\pm10.8$\\
02886 & 5119.59978 &  $78.4\pm10.0$ & 12796 & 5168.64034 &  $65.6\pm 2.6$ & 21913A & 5174.84064 &  $94.9\pm 7.2$\\
02886 & 5119.61616 &  $82.0\pm10.5$ & 12796 & 5211.66185 &  $47.4\pm 2.0$ & 23966 & 5219.70931 &  $66.7\pm 1.5$ \\
02886 & 5119.63153 &  $89.0\pm11.5$ & 12796 & 5211.67653 &  $49.2\pm 2.7$ & 23966 & 5219.72400 &  $64.3\pm 2.8$ \\
03080 & 5160.56711 &  $22.3\pm 1.3$ & 13015 & 5126.84079 & $-91.9\pm 6.0$ & 23966 & 5219.74182 &  $51.6\pm 2.0$ \\
03080 & 5160.57254 &  $21.8\pm 1.3$ & 13015 & 5126.85721 & $-92.9\pm 4.9$ & 23966 & 5219.75649 &  $51.1\pm 2.0$ \\
03080 & 5167.54764 &  $29.0\pm 1.2$ & 13015 & 5145.76540 &$-119.5\pm 9.3$ & 24406 & 5230.72715 & $119.0\pm 1.4$ \\
03080 & 5167.55306 &  $29.2\pm 0.9$ & 13015 & 5145.78181 &$-131.9\pm 3.5$ & 24406 & 5230.74356 & $119.9\pm 2.3$ \\
03080 & 5186.58826 &  $35.0\pm 1.5$ & 13015 & 5159.68154 &$-119.3\pm10.6$ & 24406 & 5236.62599 & $131.0\pm 3.0$ \\
03080 & 5186.59368 &  $34.2\pm 0.9$ & 13015 & 5159.69795 &$-134.7\pm 4.7$ & 24406 & 5236.64241 & $120.3\pm 1.5$ \\
03471 & 5129.68300 &  $30.0\pm 1.4$ & 13471 & 5205.72033 &  $46.6\pm 1.7$ & 24406 & 5241.64830 & $123.4\pm 2.0$ \\
03471 & 5129.69768 &  $30.3\pm 1.4$ & 13471 & 5205.73617 &  $41.5\pm 2.1$ & 24406 & 5241.66470 & $118.4\pm 2.8$ \\
03471 & 5130.56893 &  $27.8\pm 1.3$ & 13471 & 5221.66551 &  $34.6\pm 2.3$ & 31473 & 5227.72301 & $133.6\pm 2.2$ \\
03471 & 5130.58360 &  $27.0\pm 1.9$ & 13471 & 5221.68134 &  $30.4\pm 2.6$ & 31473 & 5227.73769 & $130.2\pm 4.3$ \\
04615 & 5130.60388 &  $50.2\pm 3.5$ & 13471 & 5227.68215 &  $53.7\pm 3.5$ & 31473 & 5228.74950 &  $90.1\pm 2.5$ \\
04615 & 5130.62030 &  $48.6\pm 2.4$ & 13471 & 5227.69800 &  $52.2\pm 2.2$ & 31473 & 5236.65995 & $118.5\pm 2.1$ \\
04615 & 5130.64061 &  $57.4\pm 2.3$ & 13532 & 5120.67703 & $112.2\pm 8.0$ & 31473 & 5236.67463 & $106.0\pm 1.1$ \\
04615 & 5130.65704 &  $50.2\pm 2.0$ & 13532 & 5120.69175 & $138.8\pm19.2$ & 31473 & 5260.60814 &  $91.9\pm 2.1$ \\
05503 & 5167.57190 &  $25.5\pm 2.2$ & 13532 & 5221.62982 & $126.3\pm 7.9$ & 31473 & 5260.62282 &  $92.1\pm19.2$ \\
05503 & 5167.58659 &  $26.9\pm 0.9$ & 13532 & 5221.64451 & $106.4\pm17.2$ & 31473 & 5261.60819 & $129.7\pm 4.6$ \\
05503 & 5168.55934 &  $24.3\pm 2.4$ & 13532 & 5228.59386 & $107.3\pm 6.2$ & 31473 & 5261.62287 & $122.2\pm 1.8$ \\
05503 & 5168.57403 &  $31.6\pm 2.8$ & 13532 & 5228.60854 & $115.7\pm11.2$ & 32785 & 5227.83769 &  $36.6\pm 4.4$ \\
05814 & 5119.65517 &  $62.2\pm 2.1$ & 13755 & 5174.69455 &  $71.8\pm 2.5$ & 32785 & 5227.84659 &  $34.0\pm 5.6$ \\
05814 & 5119.67558 &  $68.9\pm 2.3$ & 13755 & 5174.70924 &  $73.3\pm 2.3$ & 32785 & 5260.64423 &  $44.9\pm 6.8$ \\
06559 & 5202.54687 &  $69.0\pm 3.1$ & 13755 & 5174.72599 &  $82.5\pm 2.3$ & 32785 & 5260.66007 &  $25.9\pm 6.3$ \\
06559 & 5202.56328 &  $66.0\pm 1.3$ & 13755 & 5174.74067 &  $80.0\pm 2.2$ & 32785 & 5262.65007 &  $39.9\pm 5.6$ \\
06559 & 5221.54948 &  $91.5\pm 0.5$ & 14491 & 5220.64861 &  $46.5\pm 1.2$ & 32785 & 5262.66590 &  $32.5\pm 7.8$ \\
06559 & 5221.56590 &  $88.9\pm 1.9$ & 14491 & 5220.66502 &  $45.4\pm 2.7$ & 33503 & 5260.68062 & $111.8\pm 1.2$ \\
06794 & 5120.63607 &  $86.6\pm 1.4$ & 14491 & 5220.68455 &  $53.2\pm 3.0$ & 33503 & 5260.69530 & $111.3\pm 1.4$ \\
06794 & 5120.65248 &  $87.4\pm 1.8$ & 14491 & 5220.70096 &  $45.0\pm 2.2$ & 35570 & 5260.71462 &  $-0.6\pm 0.9$ \\
06794 & 5221.58891 &  $74.3\pm 2.2$ & 15882 & 5226.70469 &  $34.8\pm 1.5$ & 35570 & 5260.72930 &  $-5.9\pm 2.3$ \\
06794 & 5221.60533 &  $73.0\pm 1.2$ & 15882 & 5226.72052 &  $31.7\pm17.2$ & 50161 & 5164.55701 &   $2.4\pm17.2$ \\
07051 & 5119.69701 &  $90.5\pm 7.8$ & 15882 & 5226.74176 &  $40.9\pm 1.5$ & 50161 & 5164.57343 &   $6.9\pm16.9$ \\
07051 & 5119.71169 &  $81.1\pm 4.3$ & 15882 & 5226.75758 &  $31.0\pm 5.3$ & 51252 & 5163.55031 &  $18.3\pm 1.7$ \\
08432 & 5165.57389 &  $66.1\pm 2.8$ & 15882 & 5228.63304 &  $37.2\pm 3.1$ & 51252 & 5163.56673 &  $19.2\pm 1.6$ \\
08432 & 5165.59036 &  $63.9\pm 3.2$ & 15882 & 5228.64888 &  $37.3\pm 1.8$ & 56045 & 5186.55285 &  $19.7\pm 2.1$ \\
08432 & 5228.55214 &  $64.8\pm 2.3$ & 15882 & 5228.71117 &  $39.2\pm 2.1$ & 56045 & 5186.56926 &  $19.0\pm 1.8$ \\
08432 & 5228.56855 &  $67.8\pm 2.3$ & 15882 & 5228.72700 &  $33.9\pm 3.6$ & 56257 & 5165.53139 &  $40.6\pm 1.9$ \\
09940 & 5159.72102 &  $79.2\pm 4.4$ & 15957 & 5174.76253 & $108.0\pm 3.0$ & 56257 & 5165.54607 &  $40.7\pm 2.4$ \\
09940 & 5159.73744 &  $72.4\pm 5.7$ & 15957 & 5174.77720 & $106.9\pm 5.9$ & 56257 & 5181.55685 &  $55.9\pm 1.4$ \\
09940 & 5159.75445 &  $76.7\pm 4.6$ & 15957 & 5174.79403 & $113.7\pm 2.4$ & 56257 & 5181.57154 &  $54.7\pm 2.4$ \\
09940 & 5159.77088 &  $92.8\pm 4.9$ & 15957 & 5174.80871 & $100.3\pm 2.3$ & 56493 & 5181.59296 &  $88.6\pm 5.4$ \\
12758 & 5127.78391 &  $92.3\pm 2.0$ & 17486 & 5119.76508 & $122.5\pm 2.3$ & 56493 & 5181.60880 &  $69.6\pm11.8$ \\
12758 & 5127.79512 &  $91.1\pm 1.8$ & 17486 & 5119.77977 & $127.4\pm 2.5$ &       &            & \\

\hline
\end{tabular}\\
\end{minipage}
\end{table*}

\label{lastpage}

\end{document}